\documentclass[a4paper,11pt]{article}
\usepackage{jheppub}
\usepackage[dvipsnames]{xcolor}
\usepackage{amsfonts,amsmath,amssymb,bm}
\usepackage{graphicx}
\usepackage{epsfig,hyperref,wasysym,tikz}

\makeatletter
\def\@fpheader{\relax}
\makeatother

\protected\def\betterat{{\fontfamily{ptm}\selectfont @}}
\begingroup\lccode`~=`@ \lowercase{\endgroup\let~}\betterat
\catcode`@=\active

\title{On isomorphisms between quiver Yangians}

\author[*,\sharp,\ddagger]{Vishnu Jejjala}
\author[*]{\!\!, Dumisani Nxumalo}
\author[*,\dagger,\ddagger]{\!\!, Konstantinos Zoubos}

\affiliation[\, *]{Mandelstam Institute for Theoretical Physics \& School of Physics, University of the Witwatersrand, Johannesburg, WITS 2050, South Africa}
\affiliation[\, \sharp]{NSF AI Institute for Artificial Intelligence and Fundamental Interactions (IAIFI) and\\ Department of Physics, Northeastern University, Boston, MA 02115, USA}
\affiliation[\, \dagger]{Department of Physics, University of Pretoria, Private Bag X20, Hatfield 0028, South Africa}
\affiliation[\, \ddagger]{National Institute for Theoretical and Computational Sciences (NITheCS), Gauteng, South Africa}

\emailAdd{vishnu.jejjala@gmail.com}
\emailAdd{gudunnkomo@gmail.com}
\emailAdd{kzoubos@up.ac.za}

\newcommand\be{\begin{equation}}
\newcommand\ee{\end{equation}}
\newcommand\bea{\begin{eqnarray}}
\newcommand\eea{\end{eqnarray}}

\newcommand{\half}{\frac12}

\newcommand{\Fcal}{\mathcal{F}}

\newcommand{\Wcal}{\mathcal{W}}

\newcommand{\Ncal}{\mathcal{N}}

\newcommand{\Ecal}{\mathcal{E}}

\newcommand{\Cset}{{\,\,{{{^{_{\pmb{\mid}}}}\kern-.47em{\mathrm C}}}}}

\newcommand{\diff}{\mathrm{d}}

\newcommand{\ra}{\rightarrow}

\newcommand{\tpsi}{\tilde{\psi}}

\newcommand{\comment}[1]{}

\newcommand{\eone}{e^{(1)}_I}

\newcommand{\ethree}{e^{(3)}_I}
\newcommand{\efour}{e^{(4)}_I}
\newcommand{\fone}{f^{(1)}_I}

\newcommand{\fthree}{f^{(3)}_I}
\newcommand{\ffour}{f^{(4)}_I}

\newcommand{\psit}{\tilde{\psi}}

\newcommand{\psione}{\psi^{(1)}_I}

\newcommand{\psithree}{\psi^{(3)}_I}
\newcommand{\psifour}{\psi^{(4)}_I}

\newcommand{\tpsifour}{\tilde{\psi}^{(4)}_I}
\newcommand{\tpsitwo}{\tilde{\psi}^{(2)}_I}

\newcommand{\Fzero}{\mathbb{F}_0}
\newcommand{\pb}[1]{\{#1\}}

\abstract{Quiver Yangians are infinite-dimensional algebras capturing the BPS structure of a large class of supersymmetric models. Quiver theories related by Seiberg duality are expected to have isomorphic quiver Yangians, and this isomorphism has previously been shown for quivers corresponding to generalised conifold geometries. In this work, we present an explicit isomorphism for the two Seiberg dual phases of the $\Fzero$ quiver theory, which falls outside of the above class. Some aspects of our construction are similar to the known cases, while others appear to be specific to the $\Fzero$ quiver. In particular, the map involves square roots of operators bilinear in the fermionic fields of the mode being dualised.  
}

\begin{document}

\vspace{1cm}

\maketitle

\section{Introduction}

For a supersymmetric quantum field theory, the protected, or BPS, states are an important defining feature. Equally important is finding the algebra formed by these BPS states under fusion~\cite{Harvey:1996gc}. For the BPS sector of IIA string theory compactified on a toric Calabi--Yau threefold, the structure of this algebra was studied in~\cite{Li:2020rij}. It was called a quiver Yangian, as it generalises the usual notion of a Yangian algebra~\cite{Drinfeld88} using information read off from the quiver description of the theory. As shown in~\cite{Galakhov:2020vyb}, quiver Yangians correctly count BPS states in the $\Ncal=4$ supersymmetric quantum mechanics on the worldvolume of D-branes wrapping appropriate cycles in the Calabi--Yau manifold. See the review~\cite{Yamazaki:2022cdg} for a summary of the main features of quiver Yangians and their representations.  

 The quiver Yangians were generalised to toroidal and elliptic cases in~\cite{Noshita:2021ldl,Galakhov:2021vbo}, related to 2d and 3d theories with four supercharges, as well as to shifted quiver Yangians~\cite{Galakhov:2021xum}, allowing the construction of more general representations, and were recently extended to a much larger class of quivers~\cite{Li:2023zub}. Additional aspects of quiver Yangians for non-toric affine Dynkin diagrams were considered in~\cite{Bao:2023ece}. There are also interesting connections to integrable systems~\cite{Bao:2022fpk,Galakhov:2022uyu}. Although the study of quiver Yangians, their representations and interconnections to other algebraic structures is still in its infancy, it is clear that they represent a powerful and unifying tool in the study of supersymmetric field theory. 
 
It is well known that different quiver theories can be mapped to each other by Seiberg duality~\cite{Seiberg:1994pq}, which in this context is equivalent to toric duality~\cite{Beasley:2001zp,Feng:2002zw}. An intuitive approach to Seiberg duality is to apply urban renewal, a graphical procedure implemented at the level of the bipartite graph dual to a given quiver~\cite{Hanany:2005ve,Franco:2005rj,Franco:2005sm}, see, for instance.,~\cite{Yamazaki:2008bt} for a review. In this way the bipartite graphs corresponding to Seiberg dual phases can be seen to be related by simple operations which modify the connectivity of the graph but still preserve certain invariants~\cite{Hanany:2011bs}.

In~\cite{Li:2020rij}, it was conjectured that the quiver Yangians corresponding to Seiberg-dual phases will be isomorphic, being related to the same underlying toric geometry. To confirm this, and thus also provide an additional check of the quiver Yangian construction, it is important to construct these maps explicitly at the level of the algebra. For the case of generalised conifolds, where there are no compact four-cycles in the Calabi--Yau and the corresponding quivers are non-chiral, the structure of this map was elucidated in~\cite{Bao:2022jhy}, which worked at the level of the mode expansion. For the toroidal non-chiral case, the map was worked out in~\cite{Bao:2023kkh}, and can also be seen to follow from the isomorphism between quantum affine superalgebras for different parity sequences ~\cite{Bezerra_2020,BezerraMukhin19b} (with the construction also relying partly on the mode expansion). The work~\cite{Bao:2023kkh} also provided a preliminary discussion of the desired features of the isomorphism for toric Calabi--Yau spaces \emph{with} compact four-cycles, where the corresponding quivers are chiral. However, a concrete map for these cases is not known. As constructing the isomorphism for arbitrary quivers is challenging, it is important to consider specific examples to build up intuition which could help to understand the general structure of the isomorphism. 

In this note, we study the isomorphism of the quiver Yangians corresponding to the total space of the canonical bundle over the zeroth Hirzebruch surface $\Fzero=\mathbb{P}^1\times\mathbb{P}^1$, or $\mathbf{K}_{\Fzero}$. The associated gauge theory arises from considering a singular limit of the Calabi--Yau threefold $\mathbf{K}_{\Fzero}$. The toric description contains compact four-cycles and therefore falls outside the generalised conifold cases studied in~\cite{Bao:2022jhy}. The $\Fzero$ gauge theory has two Seiberg dual phases ~\cite{Feng:2000mi,Feng:2001xr,Feng:2002zw, Franco:2005rj}, and the simplicity of the corresponding quivers makes it a convenient setting to study the algebra isomorphisms. We provide an explicit map of the currents of the quiver Yangians corresponding to the two phases. The key element of the map is a square root of a bifermion operator, whose properties we discuss in some detail. 

We emphasise that the isomorphism we find at the level of the currents/OPEs is not expected to immediately transfer to the modes. As is well known, wall-crossing effects~\cite{Kontsevich:2008fj} can lead to markedly different representations of the algebra across stability chambers. For a treatment of $\Fzero$, see~\cite{Aganagic:2010qr}. In the quiver Yangian context, this can be captured by shifts in the mode expansion~\cite{Galakhov:2021xum} or equivalently a removal of crystal layers \cite{Galakhov:2021xum,Bao:2022oyn}. For affine $\mathfrak{gl}(1|1)$, it was shown in \cite{Galakhov:2024foa} that, even though the Seiberg-dual Yangians are trivially isomorphic at the level of the currents (since the quiver is self-dual in this case), the corresponding crystals (constructed from the modes) have a very different structure. So finding an explicit map between the currents should be thought of as only a first step in deciphering how representations transform under Seiberg duality.

After a very brief review of quiver Yangians in Section~\ref{SectionQY}, we describe the two phases of the $\Fzero$ theory in Section~\ref{SectionF0} and the main features of their associated quiver Yangians. The map between the currents is then presented in Section~\ref{SectionMap}, and various explicit checks are performed. We conclude with a discussion of open questions for future study. For easy reference, Appendix~\ref{AppendixQuadratic} lists the quadratic relations of the two phases. Appendix~\ref{AppendixCommutators} provides details of some of the required computations involving the square root-type operators appearing in the map.

\section{Quiver Yangians} \label{SectionQY}

Quiver Yangians were defined in~\cite{Li:2020rij} as a generalisation of the affine Yangian of $\mathfrak{gl}_1$~\cite{Tsymbaliuk:2014fvq}, which is an infinite-dimensional algebra defined by operators $e_n, f_n$ and $\psi_n$, with $n\geq 0$. These correspond to raising, lowering and Cartan operators, respectively. To define the quiver Yangian for an arbitrary quiver, one associates a set of such modes to each node of the quiver diagram. Consider a quiver diagram defined by a set of nodes labelled by $a,b,\ldots$, and a set of arrows between the nodes, labelled by $I,J,\ldots$. Each arrow comes with a charge assignment $h_I$, chosen such that the superpotential is neutral. Since each term in the superpotential is a closed loop $L$ in the quiver (or actually the periodic version of the quiver, which provides a clearer picture of the loops), this implies that the sum of the $h_I$ along every closed loop needs to vanish. These \emph{loop constraints} still leave some freedom in the charges, which can be fixed by imposing \emph{vertex constraints}, where the charges of the arrows coming into/out of each node are counted with positive/negative sign respectively. So, in all, the charges will be chosen to satisfy the
\be \label{LoopVertex}
\text{Loop Constraints:} \; \sum_{I\in L} h_I=0 \;\; \text{and Vertex Constraints:} \;\;\sum_{I\in a}\mathrm{sign}_a(I) h_I=0\;.
\ee
In the toric case, these constraints leave two charges unfixed, which are related to the two non-$R$ $U(1)$ symmetries of the supersymmetric quantum mechanics.

Given a quiver with charges defined as above, one can now associate a set of modes $e_n^{(a)},f_n^{(a)}$ and $\psi_n^{(a)}$ to each node of the quiver, which satisfy relations given by the arrows between the nodes. This defines the quiver Yangian. Referring to~\cite{Li:2020rij} for the mode relations, here we will only work with the quiver Yangian expressed in terms of currents formed by introducing a complex-valued spectral parameter $z$ and defining
\be
e^{(a)}(z)=\sum_{n=0}^\infty\frac{e_n^{(a)}}{z^{n+1}}\;,\;\;f^{(a)}(z)=\sum_{n=0}^\infty\frac{f_n^{(a)}}{z^{n+1}}\;\;\text{and}\;\;
\psi^{(a)}(z)=\sum_{n=-\infty}^\infty \frac{\psi_n^{(a)}}{z^{n+1}}\;,
\ee
where $a$ labels the nodes and we note that in general $\psi(z)$ now needs to contain negative modes. The currents $\psi^{(a)}(z)$ are always bosonic, while $e^{(a)}(z)$ and $f^{(a)}(z)$ are bosonic if there are arrows from node $a$ back to itself, and fermionic otherwise. One denotes by $|a|=0,1$ the bosonic/fermionic degree of each node. 

In terms of the currents, the quadratic relations of the quiver Yangian are as follows ~\cite{Li:2020rij}:
\be \label{QYall}
\begin{split}
  \psi^{(a)}(z)\psi^{(b)}(w)&=\psi^{(b)}(w)\psi^{(b)}(z)\;,\\
  \psi^{(a)}(z)e^{(b)}(w)&\simeq \varphi^{b\Rightarrow a}(u) e^{(b)}(w)\psi^{(a)}(z)\;,\\
  \psi^{(a)}(z)f^{(b)}(w)&\simeq \varphi^{b\Rightarrow a}(u)^{-1} f^{(b)}(w)\psi^{(a)}(z)\;,\\
  e^{(a)}(z)e^{(b)}(w)&\sim (-1)^{|a||b|} \varphi^{b\Rightarrow a}(u) e^{(b)}(w) e^{(a)}(z)\;,\\
  f^{(a)}(z)f^{(b)}(w)&\sim (-1)^{|a||b|} \varphi^{b\Rightarrow a}(u)^{-1} f^{(b)}(w) f^{(a)}(z)\;,\;\; \text{and}\\
  [e^{(a)}(z),f^{(b)}(w)\}&\sim-\delta^{a,b} \frac{\psi^{(a)}(z)-\psi^{(b)}(w)}{z-w}\;.
\end{split}
\ee
Here $u=z-w$, and $[\cdot,\cdot\}$ is an  anticommutator if both fields are fermionic, otherwise a commutator. To ensure that these relations correctly reproduce the mode relations when expanded for large $z,w$, one needs to add local terms of type $z^nw^m$ for $m\geq0$, denoted by $\simeq$ above, or additionally for $n\geq 0$, denoted by $\sim$ above. These terms are regular at $z=0$, $w=0$, or both, and so will generically not affect the mode expansions obtained by contour integrating around these points, apart from specific combinations of mode numbers.\footnote{In the following we will mostly write $\simeq$ for expressions with regular terms and reserve $\sim$ for proportionality.} The denominator in the $e-f$ relation is to be interpreted as a delta function $\delta^{\mathrm{rat.}}(z-w)=1/(z-w)$, which takes $\delta^{\mathrm{rat.}}(z-w)f(z)\simeq\delta^{\mathrm{rat.}}(z-w)f(w)$, i.e. up to regular terms (see e.g. \cite{Galakhov:2022uyu} for a discussion). Thus, in the following we will call the right-hand-sides of the $e-f$ relations ``delta function'' terms. 

Given the resemblance of the above relations to OPEs in 2d CFT, the currents are often called fields, a language which we will also adopt. 

The final ingredient in the definition is the bond factor $\varphi^{a\Rightarrow b}$, which for each pair of nodes is given by
\be \label{DefQYbond}
\varphi^{a\Rightarrow b}(u)=\frac{\prod_{I\in {b\rightarrow a}}(u+h_I)}{\prod_{I\in {a\rightarrow b}}(u-h_I)}\;.
\ee
That is, one multiplies by $(u+h_I)$ for each arrow from $b$ to $a$ and divides by $(u-h_I)$ for each arrow from $a$ to $b$. Clearly, if there are no arrows connecting nodes $a$ and $b$ then $\varphi^{a\Rightarrow b}=\varphi^{b\Rightarrow a}=1$ and the fields corresponding to these nodes simply commute/anticommute depending on their gradings. Exchanging the arrows, one finds that the bond factors satisfy:
\be \label{bondinverse}
\varphi^{a\Rightarrow b}(u)\varphi^{b\Rightarrow a}(-u)=1\;.
\ee
The above quadratic relations do not by themselves give a correct counting of BPS states by forming the crystals based on the (periodic extensions of the) quivers. One needs to impose additional higher-order relations, the Serre relations. We refer to~\cite{Li:2020rij} for a detailed discussion of the role of these relations. 

The quiver Yangians have been generalised to the toroidal as well as the elliptic case, where the rational functions appearing in the bond factors are replaced by trigonometric or theta functions, respectively, and one also has two bosonic currents $\psi_\pm(z)$ for each node~\cite{Galakhov:2021vbo,Bao:2023kkh}. We will not consider these generalised algebras in this work, however we expect that our results can be straightforwardly extended to those cases.

\section{The two phases of $\Fzero$} \label{SectionF0}

For our discussion it will be useful to recall that toric quivers can alternatively be represented as bipartite graphs, which are dual in the sense that vertices in the quiver are faces on the bipartite graph, while the arrows between the nodes are the edges of the graph. The vertices of the bipartite graph are coloured black or white to indicate the sign of the superpotential. The direction of the arrows is read off by the alternation of black and white vertices in the graph: Going clockwise around a face, a \begin{tikzpicture}[scale=0.5] \draw[-](0,4)--(1,4);\draw[fill=white] (0,4) circle (1ex);\draw[fill=black] (1,4) circle (1ex); \end{tikzpicture} edge indicates an outgoing arrow while a 
\begin{tikzpicture}[scale=0.5] \draw[-](0,4)--(1,4);\draw[fill=black] (0,4) circle (1ex);\draw[fill=white] (1,4) circle (1ex); \end{tikzpicture} edge indicates an incoming arrow. We refer to, e.g.,~\cite{Yamazaki:2008bt} for further aspects of bipartite graphs and their relation to string constructions and supersymmetric gauge theory.  

\begin{figure}[ht]
 \begin{center}
      \begin{tikzpicture}[scale=0.5,baseline=1cm]
        \draw[-,blue,thick] (0,0)--(4,0)--(4,4)--(0,4)--(0,0);
        \draw[-,blue] (-0.5,0)--(0,0);\draw[-,blue] (-0.35,-0.35)--(0,0);\draw[-,blue] (0,-0.5)--(0,0);
        \draw[-,blue] (-0.5,4)--(0,4);\draw[-,blue] (-0.35,4.35)--(0,4);\draw[-,blue] (0,4.5)--(0,4);
        \draw[-,blue] (4.5,0)--(4,0);\draw[-,blue] (4.35,-0.35)--(4,0);\draw[-,blue] (4,-0.5)--(4,0);
        \draw[-,blue] (4.5,4)--(4,4);\draw[-,blue] (4.35,4.35)--(4,4);\draw[-,blue] (4,4.5)--(4,4);
         
        \draw[fill=black] (0,0) circle (1ex);
        \draw[fill=white] (0,4) circle (1ex);
        \draw[fill=black] (4,4) circle (1ex);
        \draw[fill=white] (4,0) circle (1ex);
        \node at (8,2) {$\Leftrightarrow$};
      \end{tikzpicture}\hspace{1.5cm}
      \begin{tikzpicture}[scale=0.5,baseline=1cm]
        \draw[-,blue,thick] (1,1)--(3,1)--(3,3)--(1,3)--(1,1);
        \draw[-,blue] (-0.5,0)--(0,0);\draw[-,blue] (-0.35,-0.35)--(0,0);\draw[-,blue] (0,-0.5)--(0,0);
        \draw[-,blue] (-0.5,4)--(0,4);\draw[-,blue] (-0.35,4.35)--(0,4);\draw[-,blue] (0,4.5)--(0,4);
        \draw[-,blue] (4.5,0)--(4,0);\draw[-,blue] (4.35,-0.35)--(4,0);\draw[-,blue] (4,-0.5)--(4,0);
        \draw[-,blue] (4.5,4)--(4,4);\draw[-,blue] (4.35,4.35)--(4,4);\draw[-,blue] (4,4.5)--(4,4);

        \draw[-,blue,thick] (0,0)--(1,1);\draw[-,blue,thick] (4,0)--(3,1);\draw[-,blue,thick] (4,4)--(3,3);\draw[-,blue,thick] (0,4)--(1,3);
        \draw[fill=black] (0,0) circle (1ex);
        \draw[fill=white] (0,4) circle (1ex);
        \draw[fill=black] (4,4) circle (1ex);
        \draw[fill=white] (4,0) circle (1ex);
         \draw[fill=white] (1,1) circle (1ex);
        \draw[fill=black] (3,1) circle (1ex);
        \draw[fill=white] (3,3) circle (1ex);
        \draw[fill=black] (1,3) circle (1ex);
      \end{tikzpicture}
\caption{The square move/urban renewal transformation implementing Seiberg duality at the level of the bipartite graph.}\label{FigureUR}
      \end{center}
\end{figure}
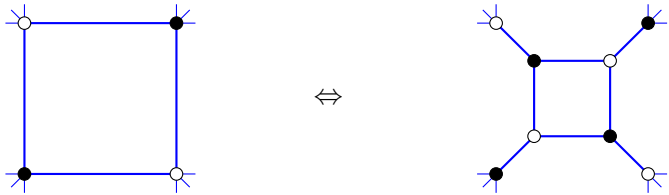

At the level of the bipartite graph, Seiberg duality is implemented through a process called \emph{urban renewal}~\cite{Franco:2005rj,Franco:2005sm}. It is a local graph operation which changes the connectivity of a given 4-sided face in the graph, as depicted in Figure~\ref{FigureUR}. After applying some simple rules (such as integrating out massive fields corresponding to vertices with only two lines), one obtains a new bipartite graph depicting the result of Seiberg dualising the corresponding gauge group. Urban renewal can result in a graph equivalent to the original one, in which case the theory has only one phase, but it can also lead to an inequivalent graph, i.e., a different phase of the theory. Dualising the same face twice brings us back to the original phase, while multiple Seiberg phases can be obtained by dualising different faces in sequence.  

Let us now focus on the quivers corresponding to the cone over the zeroth Hirzebruch surface, $\Fzero = \mathbb{P}^1\times \mathbb{P}^1$. (The total space of the canonical bundle $\mathbf{K}_{\Fzero}$, a smooth, non-compact Calabi--Yau threefold, is the crepant resolution of this cone.) The worldvolume gauge theory on a D$3$-brane at the tip of the cone is well known~\cite{Feng:2000mi,Feng:2001xr,Feng:2002zw, Franco:2005rj} to admit two phases, related by Seiberg/toric duality. The quiver Yangians for the two corresponding quiver theories are discussed in detail in~\cite{Li:2020rij}, and our conventions below are the same as in that work.   

Unlike the  generalised conifold case, for $\Fzero$ all four nodes are fermionic (i.e., there are no lines leaving and returning to the same nodes) both before and after the duality. In the generalised conifold case, the adjacent nodes to the one being dualised switch grading, which is captured by a quadratic map for the $e^{(a)}$ fields of the type $e^{(a)}_{II}\sim e^{(a)}_{I}e^{(F)}_{I}$, where $F$ denotes the dualised node (which is always fermionic), and $a=F\pm 1$~\cite{Bao:2022jhy}. A similar mapping applies to the $f$ fields. Clearly, for $\Fzero$ the structure of the map will have to be different, and in particular we expect the map to contain bosonic combinations of $e^{(F)}_I$.  

For concreteness, as in~\cite{Li:2020rij}, in the following we will perform Seiberg duality on node 4. Hence the adjacent nodes will be 1 and 3. Of course, dualising any other node is equivalent by appropriate shifts in the node numbers.

\subsection{Quiver Yangian for phase I}

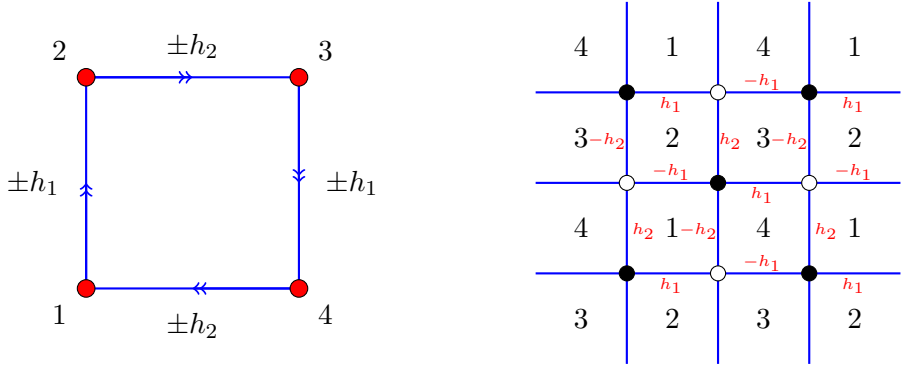
\begin{figure}[ht]
 \begin{center}
      \begin{tikzpicture}[scale=0.7,baseline=-1.5cm]
 
\draw[->,blue,thick] (0,0)--(4,0)--(4,4)--(0,4)--(0,0);
\draw[->>,blue,thick] (0,0)--(0,2);\draw[->>,blue,thick] (0,4)--(2,4);\draw[->>,blue,thick] (4,4)--(4,2);\draw[->>,blue,thick] (4,0)--(2,0);
\draw[fill=red] (0,0) circle (1ex);\draw[fill=red] (4,0) circle (1ex);\draw[fill=red] (4,4) circle (1ex);
\draw[fill=red] (0,4) circle (1ex);

\node at (-0.5,-0.5) {$1$};\node at (-0.5,4.5) {$2$};\node at (4.5,4.5) {$3$};\node at (4.5,-0.5) {$4$};
\node at (-1,2) {$\pm h_1$};\node at (2,4.6) {$\pm h_2$};
\node at (5,2) {$\pm h_1$};\node at (2,-0.7) {$\pm h_2$};
      \end{tikzpicture}\hspace{1.8cm}
      \begin{tikzpicture}[scale=0.6,baseline=-0.5cm]
 
       \draw[-,blue,thick] (0,2)--(8,2);
       \draw[-,blue,thick] (0,4)--(8,4);
       \draw[-,blue,thick] (0,6)--(8,6);
       \draw[-,blue,thick] (2,0)--(2,8);
       \draw[-,blue,thick] (4,0)--(4,8);
       \draw[-,blue,thick] (6,0)--(6,8);
       \draw[fill=black] (2,2) circle (1ex);\draw[fill=white] (4,2) circle (1ex);\draw[fill=black] (6,2) circle (1ex);
       \draw[fill=white] (2,4) circle (1ex);\draw[fill=black] (4,4) circle (1ex);\draw[fill=white] (6,4) circle (1ex);
       \draw[fill=black] (2,6) circle (1ex);\draw[fill=white] (4,6) circle (1ex);\draw[fill=black] (6,6) circle (1ex);
       \node at (3,3) {$1$};\node at (3,5) {$2$};\node at (5,5) {$3$};\node at (5,3) {$4$};
       \node at (1,3) {$4$};\node at (7,3) {$1$};\node at (1,5) {$3$};\node at (7,5) {$2$};
       \node at (1,1) {$3$};\node at (3,1) {$2$};\node at (5,1) {$3$};\node at (7,1) {$2$};
       \node at (1,7) {$4$};\node at (3,7) {$1$};\node at (5,7) {$4$};\node at (7,7) {$1$};
       \node[red] at (3,1.7) {${}^{{}_{h_1}}$};\node[red] at (5,2.2) {${}^{{}_{-h_1}}$};\node[red] at (7,1.7) {${}^{{}_{h_1}}$};
       \node[red] at (2.4,2.9) {${}^{{}_{h_2}}$}; \node[red] at (3.6,2.9) {${}^{{}_{-h_2}}$};\node[red] at (6.4,2.9) {${}^{{}_{h_2}}$};
       \node[red] at (3,4.2) {${}^{{}_{-h_1}}$};\node[red] at (5,3.7) {${}^{{}_{h_1}}$};\node[red] at (7,4.2) {${}^{{}_{-h_1}}$};
       \node[red] at (1.6,4.9) {${}^{{}_{-h_2}}$}; \node[red] at (4.3,4.9) {${}^{{}_{h_2}}$};\node[red] at (5.6,4.9) {${}^{{}_{-h_2}}$};
       \node[red] at (3,5.7) {${}^{{}_{h_1}}$};\node[red] at (5,6.2) {${}^{{}_{-h_1}}$};\node[red] at (7,5.7) {${}^{{}_{h_1}}$};
     
      \end{tikzpicture}
\caption{The quiver and bipartite diagram for phase I of $\Fzero$. We have indicated the charges satisfying the loop and vertex constraints~(\ref{LoopVertex}). To better visualise the direction of the arrows in the bipartite graph, we have placed the charges on the side of each edge where the arrow is pointing. } \label{FigPhaseI}
 \end{center}
\end{figure}

The first phase of $\Fzero$ is illustrated in Figure~\ref{FigPhaseI}, which presents the quiver as well as the corresponding bipartite diagram.
We have also indicated the charges solving the vertex and loop constraints. The charge assignment is chosen to agree with~\cite{Li:2020rij}. Note that instead of assigning the charges via the periodic quiver, we use the bipartite graph, which contains the same information. In particular, the loop constraints are read off by adding the charges around each vertex in the direction of the arrows (clockwise for black nodes and anti-clockwise for white nodes), while the vertex constraints are read off by going around each face and adding the charges with (+/-) signs depending whether the arrows are incoming or outgoing. 

From the definition~(\ref{DefQYbond}) we see that there are two independent bond factors,
\be \label{phi12}
\begin{split}
  \varphi^{2\Rightarrow 1}(u)&=\varphi^{4\Rightarrow 3}(u)=\varphi_1(u)=(u+h_1)(u-h_1)\;,\\
  \varphi^{3\Rightarrow 2}(u)&=\varphi^{1\Rightarrow 4}(u)=\varphi_2(u)=(u+h_2)(u-h_2)\;,
\end{split}
\ee
with the remaining ones obtained via~(\ref{bondinverse}). For instance, $\varphi^{1\Rightarrow2}(u)=\varphi_1(u)^{-1}$, where we took into account that the bond factors in this case are even in $u$. 

An important feature of these bond factors is that one does not have an equal number of numerator and denominator factors, as was the case for the generalised conifold quivers. In that setting, the mode expansion of $\psi(z)$ can be truncated to the modes with $n\geq -1$ \cite{Li:2020rij}, which considerably simplifies working with the modes, but this is not the case for chiral quivers such as $\Fzero$. In the following we will work purely with the fields rather than the mode expansion. 

The full set of quadratic relations of phase I is listed in Appendix~\ref{AppPhaseI}. We note in particular that since there are no arrows/edges between nodes/faces 3 and 1, the associated fields simply (anti)commute:
\be
\eone(z)\ethree(w)\sim -\ethree(w)\eone(z) \;\;,\;\; \fone(z)\fthree(w)\sim -\fthree(w)\fone(z) \;, 
\ee
and similarly for the $\psione-\ethree$, $\psione-\fthree$, $\psithree-\eone$ and $\psithree-\fone$ relations.

\subsection{Quiver Yangian for phase II}

\begin{figure}[ht]
\begin{center}
      \begin{tikzpicture}[scale=0.7,baseline=-1.2cm]
 
        \draw[->,blue,thick] (0,0)--(4,0)--(4,4)--(0,4)--(0,0);
        \draw[->,blue,thick] (0,0)--(4,4);
        \draw[->>,blue,thick] (0,0)--(0,2);\draw[->>,blue,thick] (0,4)--(2,4);\draw[->>,blue,thick] (4,0)--(4,2);\draw[->>,blue,thick] (0,0)--(2,0);
        \draw[->>>>,blue,thick] (4,4)--(1.8,1.8);
        \draw[fill=red] (0,0) circle (1ex);\draw[fill=red] (4,0) circle (1ex);\draw[fill=red] (4,4) circle (1ex);
\draw[fill=red] (0,4) circle (1ex);
\node at (-0.5,-0.5) {$1$};\node at (-0.5,4.5) {$2$};\node at (4.5,4.5) {$3$};\node at (4.5,-0.5) {$4$};
\node at (-1,2) {$\pm h_1$};\node at (2,4.7) {$\pm h_2$};
\node at (5,2) {$\mp h_1$};\node at (2,-0.7) {$\mp h_2$};
\node at (2.5,1.4) {${}_{\pm h_1\!\pm \!h_2}$};
      \end{tikzpicture}\hspace{2cm}
      \begin{tikzpicture}[scale=0.5,baseline=0.6cm]
 
        \draw[-,blue,thick] (4,6)--(4,4)--(5.5,2.5)--(7.5,2.5)--(9,4)--(9,6)--(7.5,7.5)--(5.5,7.5)--(4,6);
        \draw[-,blue,thick] (7.5,9.5)--(7.5,7.5)--(9,6)--(11,6)--(12.5,7.5)--(12.5,9.5)--(11,11)--(9,11)--(7.5,9.5);
        \draw[-,blue,thick] (5.5,7.5)--(5.5,9.5)--(7.5,9.5);\draw[-,blue,thick] (5.5,9.5)--(4,11);
        \draw[-,blue,thick] (9,4)--(11,4)--(11,6);\draw[-,blue,thick] (11,4)--(12.5,2.5);
\draw[-,blue,thick] (5.5,2.5)--(5.5,1.5);\draw[-,blue,thick] (7.5,2.5)--(7.5,1.5);\draw[-,blue,thick] (12.5,1.5)--(12.5,2.5)--(13.5,2.5);
\draw[-,blue,thick] (4,4)--(3,4);\draw[-,blue,thick] (4,6)--(3,6);\draw[-,blue,thick] (3,11)--(4,11)--(4,12);
\draw[-,blue,thick] (9,11)--(9,12);\draw[-,blue,thick] (11,11)--(11,12);\draw[-,blue,thick] (12.5,9.5)--(13.5,9.5);\draw[-,blue,thick] (12.5,7.5)--(13.5,7.5);

\draw[fill=black] (5.5,9.5) circle (1ex);\draw[fill=white] (7.5,9.5) circle (1ex);
\draw[fill=white] (5.5,7.5) circle (1ex);\draw[fill=black] (7.5,7.5) circle (1ex);
\draw[fill=white] (9,6) circle (1ex);\draw[fill=black] (11,6) circle (1ex);
\draw[fill=black] (9,4) circle (1ex);\draw[fill=white] (11,4) circle (1ex);
\draw[fill=white] (4,11) circle (1ex);\draw[fill=black] (9,11) circle (1ex);\draw[fill=white] (11,11) circle (1ex);
\draw[fill=black] (4,6) circle (1ex);\draw[fill=white] (4,4) circle (1ex);
\draw[fill=black] (5.5,2.5) circle (1ex);\draw[fill=white] (7.5,2.5) circle (1ex);\draw[fill=black] (12.5,2.5) circle (1ex);
\draw[fill=black] (12.5,9.5) circle (1ex);\draw[fill=white] (12.5,7.5) circle (1ex);

\node at (6.5,5) {$1$};\node at (6.5,8.5) {$2$};\node at (10,8.5) {$3$};\node at (10,5) {$4$};
\node at (3,8.5) {$3$};\node at (3,5) {$4$};\node at (13.5,5) {$1$};
\node at (6.5,12) {$1$};\node at (10,12) {$4$};\node at (13.5,12) {$1$};\node at (13.5,8.5) {$2$};

\node[red]  at (5,8.5) {${}^{{}_{-h_2}}$};\node[red] at (8,8.5) {${}^{{}_{h_2}}$};\node[red] at (12,8.5) {${}^{{}_{-h_2}}$};
\node[red]  at (6.5,7.7) {${}^{{}_{-h_1}}$};\node[red] at (6.5,9.1) {${}^{{}_{h_1}}$};\node[red] at (10,10.6) {${}^{{}_{h_1}}$};
\node[red]  at (9.4,5) {${}^{{}_{h_2}}$};\node[red] at (10,6.2) {${}^{{}_{-h_1}}$};
\node[red]  at (10.6,5) {${}^{{}_{-h_2}}$};\node[red]  at (3.6,5) {${}^{{}_{-h_2}}$};
\node[red] at (10,3.6) {${}^{{}_{h_1}}$};\node[red] at (6.5,2.1) {${}^{{}_{h_1}}$};

\node[red,rotate=45] at (7.8,10.5) {${}^{{}_{-h_1-h_2}}$};\node[red,rotate=45] at (7.8,3.5) {${}^{{}_{-h_1-h_2}}$};
\node[red,rotate=-45] at (12,10.5) {${}^{{}_{-h_1+h_2}}$};\node[red,rotate=-45] at (12,3.5) {${}^{{}_{-h_1+h_2}}$};
\node[red,rotate=-45] at (7.8,6.5) {${}^{{}_{h_1-h_2}}$};\node[red,rotate=-45] at (5,3.5) {${}^{{}_{-h_1+h_2}}$};
\node[red,rotate=45] at (12,6.5) {${}^{{}_{h_1+h_2}}$};\node[red,rotate=45] at (5,6.5) {${}^{{}_{h_1+h_2}}$};

      \end{tikzpicture}
\caption{The quiver and bipartite diagram for phase II of $\Fzero$. Note the exchange of white and black nodes for face 4 (the one being dualised), corresponding to the reversal of all the arrows involving that node in the quiver. We again indicate the charges satisfying~(\ref{LoopVertex}), with the charge assignments for node/face 2 having been kept the same as for phase I.} \label{FigPhaseII}
\end{center}
\end{figure}
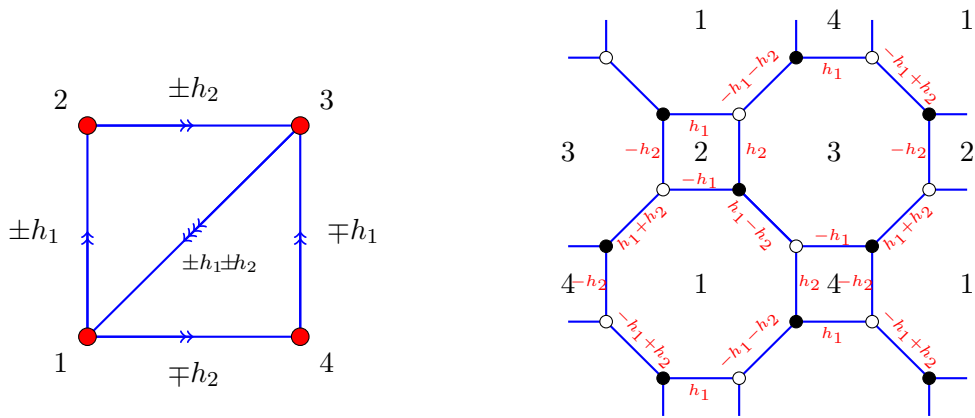

As we can see in Figure~\ref{FigPhaseII}, dualising node 4 results in a change in orientation of the $4-1$ and $3-4$ arrows, as well as the linking of nodes $1-3$ by a quadruple arrow. Therefore, recalling~(\ref{DefQYbond}), the bond factors between nodes $4-1$ and $3-4$ will be inverses of those in phase I. Furthermore, a new bond factor will emerge between phases $3-1$:\footnote{This corrects a typo in~\cite{Li:2020rij} (arXiv version 3). We thank W.\ Li for correspondence on this issue.}

\be \label{phi3}
\varphi^{1\Rightarrow3}(u)=\varphi_3(u)=(u+h_1+h_2)(u-h_1-h_2)(u+h_1-h_2)(u-h_1+h_2)\;.
\ee
In writing this factor we have already imposed the loop and vertex constraints, making sure that the assignment of charges for the arrows involving node 2 is the same as in phase I.\footnote{This is not always possible, as in some cases (such as the del Pezzo quivers) the change in connectivity after urban renewal can lead to new charges for arrows involving non-adjacent nodes.}

The list of quadratic relations for phase II can be found in Appendix~\ref{AppPhaseII}. An important difference to phase I is that now the OPEs between fields of nodes/faces 3 and 1 acquire a non-trivial bond factor:
\be
e^{(3)}_{II}(z)e^{(1)}_{II}(w)\sim- \varphi_3(u) e^{(1)}_{II}(w) e^{(3)}_{II}(z)\;,\;f^{(3)}_{II}(z)f^{(1)}_{II}(w)\sim-\varphi_3(u)^{-1} f^{(1)}_{II}(w) f^{(3)}_{II}(z)\;,\; \text{etc.}
\ee
In the following section we will turn to the question of how to obtain fields satisfying the OPEs of phase II from those of phase I.

\section{Mapping the fields} \label{SectionMap}

Our goal in this section is to express the fields of phase II of $\Fzero$ in terms of those of phase I, in a way that reproduces all the phase II OPEs. As all the phase I OPEs involve either $\varphi_1$ or $\varphi_2$ bond factors, the $\varphi_3$ bond factor needs to be constructed via OPEs involving just these two bond factors. Although alternative constructions could be possible, our approach is to generate $\varphi_3$ as a product of $\varphi_{1,2}$ at shifted values of the spectral parameter. In particular, from (\ref{phi12}) and (\ref{phi3}) we observe that
\be \label{phi3separate}
  \varphi_3(u)=\varphi_1(u+h_2)\varphi_1(u-h_2)=\varphi_2(u+h_1)\varphi_2(u-h_1)\;.
\ee
As we will see, it will be more convenient to combine these relations to express
\be \label{phi3map}
\varphi_3(u)=\left(\varphi_1(u+h_2)\varphi_1(u-h_2)\varphi_2(u+h_1)\varphi_2(u-h_1)\right)^\half\;.
\ee
Our map will make use of appropriate shifts in the spectral parameters at nodes $1$ and $3$, in order to reproduce this relation.\footnote{This relation of course simplifies at special values of the charges, such as $h_1=\pm h_2$, but we are interested in keeping the charges generic.}

Given the local nature of urban renewal, we expect that the only nodes transformed under Seiberg duality are the dualised node itself and the two adjacent nodes. As our convention is to dualise node 4, we need to provide maps for all the currents associated to nodes 4, 1 and 3. The currents associated to node 2 will be unaffected by the duality: $e^{(2)}_{II}(z)=e^{(2)}_{I}(z)$.

\subsection{Dualised node}

The map for the dualised node, which is node 4 in our conventions (see Figure~\ref{FigPhaseII}) is taken to be:
\be \label{Map4}
\begin{split}
e^{(4)}_{II}(z)&=f^{(4)}_I(z)\tpsi_I^{(4)}(z)\;,\\
f^{(4)}_{II}(z)&=-e^{(4)}_I(z)\tpsi_I^{(4)}(z)\;,\\
\psi^{(4)}_{II}(z)&=\tpsi^{(4)}_I(z)\;.
\end{split}
\ee
Here the notation $\tpsi_I^{(4)}(z)$  indicates the inverse of $\psi_I^{(4)}(z)$, such that $\tpsi_I^{(4)}(z)\psi_I^{(4)}(z)=1$.

It is easy to check that the quadratic relations of node 4 are self-dual under this map. The only nontrivial case is the $e-f$ relation:
\be
\begin{split}
  e_{II}^{(4)}(z)f_{II}^{(4)}(w)&=\left(f_I^{(4)}(z)\tpsi_I^{(4)}(z)\right)\left(-e_I^{(4)}(w)\tpsi_I^{(4)}(w)\right)=-f_I^{(4)}(z)e_I^{(4)}(w)\tpsi_I^{(4)}(z)\tpsi_I^{(4)}(w)\\
  &=\left(+e_I^{(4)}(w)f_I^{(4)}(z)+\frac{\psi_I^{(4)}(z)-\psi_I^{(4)}(w)}{z-w}\right)\tpsi_I^{(4)}(z)\tpsi_I^{(4)}(w)\\
&=-\left(-e_I^{(4)}(w)\tpsi_I^{(4)}(w)\right)\left(f_I^{(4)}(z)\tpsi_I^{(4)}(z)\right)+\frac{\tpsi_I^{(4)}(w)-\tpsi_I^{(4)}(z)}{z-w}\\
&=-f_{II}^{(4)}(w)e_{II}^{(4)}(z)-\frac{\psi_{II}^{(4)}(z)-\psi_{II}^{(4)}(w)}{z-w}\;.
\end{split}
\ee
This map follows the general pattern outlined in~\cite{Bao:2022jhy} and~\cite{Bao:2023kkh}: The $e$ and $f$ currents are exchanged while $\psi$ is mapped to its inverse. Since, as we will see, the maps for the adjacent nodes are linear in the currents associated to those nodes (e.g., $e^{(3)}_{II}$ will be linear in $e^{(3)}_I$), this exchange guarantees the inversion of the $3-4$ and $1-4$ bond factors in phase II as compared to phase I. 

 Since the same map appeared in~\cite{Bao:2022jhy,Bao:2023kkh} which studied the generalised conifold case, while we are considering $\Fzero$, it is natural to expect that~(\ref{Map4}) is the generic first step in the dualisation of any quiver Yangian.

\subsection{Adjacent nodes} \label{MapAdjacent}

Now let us consider the maps for nodes 1 and 3, which are adjacent to the node which is being dualised. As one can see by comparing~(\ref{PhaseI34}) with~(\ref{PhaseII34}) and~(\ref{PhaseI41}) with~(\ref{PhaseII41}), all the bond factors in these relations are simply inverted.  To achieve this inversion it is sufficient to choose the adjacent currents of phase II to be linear in those of phase I:
\be \label{Map13initial}
e^{(1,3)}_{II}(z)=e_{I}^{(1,3)}(z) \left[\cdots\right]\;,\;f^{(1,3)}_{II}(z)=f_{I}^{(1,3)}(z) \left[\cdots\right]\;,\;\psi^{(1,3)}_{II}(z)=\psi_{I}^{(1,3)}(z) \left[\cdots\right]\;,
\ee
where $[\cdots]$ denotes any additional combination of fields, which needs to be bosonic to avoid changing the gradings of the node 1 and 3 fields, as well as to commute, up to regular terms, with the fields of node 4 and 2.

As an example we check:
\be
\begin{split}
  e_{II}^{(4)}(z)~ e_{II}^{(1)}(w)&= \left(\ffour(z)\tpsifour(z)\right) \left(\eone(w)\left[\cdots\right]\right)
  \simeq\ffour(z)\left(\varphi_2(u)^{-1}\eone(w)\tpsifour(z)\right)\left[\cdots\right]\\
&\simeq-\varphi_2(u)^{-1}\left(\eone(w)\left[\cdots\right]\right)\left(\ffour(z)\tpsifour(z)\right)\simeq-\varphi_2(u)^{-1} e_{II}^{(1)}(w)e_{II}^{(4)}(z)\;,
\end{split}
\ee
where $u=z-w$ as always. We note that the needed minus comes from the  $e^{(1)}-f^{(4)}$ anticommutation relation, while the bond factor comes from the $e^{(1)}-\tilde{\psi}^{(4)}$ relation (which can be found by multiplying the first relation in~(\ref{PhaseI41}) by $\tilde{\psi}^{(4)}$ from the left and right).

Similarly we can check the $e^{(4)}-f^{(1)}$ relation:
\be
\begin{split}
e^{(4)}_{II}(z)~ f^{(1)}_{II}(w)&=\left(\ffour(z)\tpsifour(z)\right)\left(\fone(w) \left[\cdots\right]\right)\\
&\simeq(-\varphi_2^{-1}(u))(\varphi_2(u))\left(\fone(w) \left[\cdots\right]\right)\left(\ffour(z)\tpsifour(z)\right)\simeq- f^{(1)}_{II}(w) ~ e^{(4)}_{II}(z)\;,
\end{split}
\ee
where we see that the two bond factors from the $\tpsi^{(4)}-f^{(1)}$ and $f^{(4)}-f^{(1)}$ relations cancel out. The remaining required relations (such as $e^{(3)}-e^{(4)}$, $e^{(3)}-\psi^{(4)}$ etc.) can be seen to hold as well. As the fields of node 2 are unaffected by the duality on node 4, the $2-3$ and $2-1$ relations will stay the same, of course as long as the $[\cdots]$ commutes with the fields of node 2.

So the main challenge in finding the map for the adjacent nodes is to obtain the correct $1-3$ relations, which involve the new $\varphi_3$ bond factor (compare~(\ref{PhaseI31}) with~(\ref{PhaseII31})). To achieve this, and fix the $[\cdots]$ factors above, we first note (see~(\ref{PhaseI34})) that
\be \label{e3E}
\ethree(z) \left[\efour(w\!+\!h_2)\efour(w\!-\!h_2)\right]\simeq\varphi_1(u\!-\!h_2)\varphi_1(u\!+\!h_2) \left[\efour(w\!+\!h_2)\efour(w\!-\!h_2)\right] \ethree(z)\;,
\ee
while from~(\ref{PhaseI41}) we have
\be \label{e1E}
\left[\efour(z\!+\!h_1) \efour(z\!-\!h_1)\right] \eone(w) \simeq\varphi_2(u\!+\!h_1)\varphi_2(u\!-\!h_1) \eone(w) \left[\efour(z\!+\!h_1)\efour(z\!-\!h_1)\right]\;.
\ee
Motivated by this, we define the following bosonic operators, which are bilinear in the fermionic currents of node 4:
\be
\Ecal_{h_1}(z)=\efour(z+h_1) \efour(z-h_1) \;,\;\; \text{and}\;\; \Ecal_{h_2}(z)=\efour(z+h_2) \efour(z-h_2)\;.
\ee
In order to obtain the desired bond factors we will need to consider the \emph{square roots} of the above operators, $\Ecal^\half_{h_i}(z)$. This is not unlike the recently-proposed root-$T\bar{T}$ operator~\cite{Ferko:2022cix}, appearing in the context of marginal deformations of 2d CFT, which is also a square root of a product of two currents. Like arbitrary powers of operators, it can be straightforwardly defined through the Schwinger parametrisation (see~\cite{Hadasz:2024pew} for a detailed discussion in the root-$T\bar{T}$ context). We will adopt a similar approach here, and define 
\be
\Ecal^\half_{h_i}(z)=\frac{1}{4\sqrt{\pi}}\oint_{C}\diff s\; ~s^{-\frac{3}{2}} e^{-s \Ecal_{h_i}(z)}\;,
\ee
where the contour is the Hankel contour, taken in the positive direction around the positive real axis. Of course, unlike the root-$T\bar{T}$ case, our $\Ecal_{h_i}$ is a fermion bilinear, so one would have to carefully define the exponential through a suitable point-splitting prescription. Leaving this for future study, we treat the above definition formally.\footnote{The affine Yangian of $\mathfrak{gl}_1$ and its supersymmetric generalisation have special points where a free-field construction is known~\cite{Prochazka:2015deb,Gaberdiel:2017dbk,Gaberdiel:2017hcn}. One would expect the $\Fzero$ Yangian to have analogous points, where the fermionic currents $e$ and $f$ can themselves be expressed as boson-fermion bilinears. At such points (which however correspond to non-generic values of the $h_i$), $\Ecal^\half_{h_i}$ might have a natural description in terms of the underlying free fermions/bosons.} Some useful formulas for computing with the root-$\Ecal$ operator can be found in Appendix~\ref{AppendixCommutators}. 

Given the homogeneous nature of the $e^{(i)}-e^{(4)}$ relations, when passing an $e^{(i)}$ from left to right of $\Ecal^{\half}_{h_i}$ or vice versa, one first expands the exponential, and notices that commuting with a $k^{\mathrm{th}}$-order term gives the other ordering times a $k^{\mathrm{th}}$ power of the bond factors, so the result can be easily re-exponentiated. In this way we can verify the relations
\be
\begin{split}
\ethree(z)\Ecal^\half_{h_2}(w)&\simeq (\varphi_1(u-h_2)\varphi_1(u+h_2))^\half \Ecal^\half_{h_2}(w)\ethree(z)\;\; \text{and}\\ \;\;
\Ecal^\half_{h_1}(z)\eone(w)&\simeq (\varphi_2(u+h_1)\varphi_2(u-h_1))^\half \eone(w)\Ecal^\half_{h_1}(z)\;.
\end{split}\ee
Acting twice on the square root operators, one of course recovers~(\ref{e3E}) and~(\ref{e1E}). Similarly we can define the operators
\be
\Fcal_{h_1}(z)=\ffour(z+h_1) \ffour(z-h_1) \;,\;\; \text{and}\;\; \Fcal_{h_2}(z)=\ffour(z+h_2) \ffour(z-h_2)\;,
\ee
which give 
\be
\begin{split}\fthree(z)\Fcal^\half_{h_2}(w)&\simeq (\varphi_1(u-h_2)\varphi_1(u+h_2))^{-\half} \Fcal^\half_{h_2}(w)\fthree(z)\;\; \text{and}\\ \;\;
  \Fcal^\half_{h_1}(z)\fone(w)&\simeq (\varphi_2(u+h_1)\varphi_2(u-h_1))^{-\half} \fone(w)\Fcal^\half_{h_1}(z)\;.
\end{split}
\ee
If we now define the map from phase I to phase II fields to be
\be\label{Map13ef}
\boxed{\begin{split}
e_{II}^{(1)}(z)&= \Ecal^\half_{h_2}(z)\eone(z) \;\;,\;\; e_{II}^{(3)}(z)=\ethree(z) \Ecal^\half_{h_1}(z)\;,\\
f_{II}^{(1)}(z)&=\fone(z) \Fcal^\half_{h_2}(z) \;\;,\;\; f_{II}^{(3)}(z)= \Fcal^\half_{h_1}(z) \fthree(z)\;,
\end{split}}\ee
we can easily verify that
\be
\begin{split}
  e_{II}^{(3)}(z) e_{II}^{(1)}(w)&=\left(\ethree(z) \Ecal^\half_{h_1}(z)\right)\left( \eone(w) \Ecal^\half_{h_2}(w)\right)\\
  &\simeq-\varphi_3(u) \left(\eone(w) \Ecal^\half_{h_2}(w)\right)\left( \ethree(z) \Ecal^\half_{h_1}(z)\right)
\simeq-\varphi_3(u) e_{II}^{(1)}(w) e_{II}^{(3)}(z)\;,
\end{split}
\ee
where we used~(\ref{phi3map}) and the fact that $\Ecal^\half(z)$ commutes with $\Ecal^\half(w)$. Similarly we can check
\be
f_{II}^{(3)}(z) f_{II}^{(1)}(w)\simeq-\varphi_3^{-1}(u) f_{II}^{(1)}(w) f_{II}^{(3)}(z)\;.
\ee
We emphasise that if we had not taken the square roots of our operators, the above relations would have given the squares of the required bond factors. 

Note that the order in which the fields are multiplied in~(\ref{Map13ef}) is important, since the different orderings are related as (for example):
\be
\begin{split}
\Ecal^\half_{h_2}(z) \eone(z') &\simeq\sqrt{\varphi_2(z-z'+h_2)\varphi_2(z-z'-h_2)}\eone(z') \Ecal^\half_{h_2}(z)\\
&=(z-z')\sqrt{(z-z')^2-4h_2^2}~ \eone(z') \Ecal^\half_{h_2}(z)\;.
\end{split}
\ee
Defining the operators in~(\ref{Map13ef}) by point-splitting and taking the coincident limit $z'\ra z$, we see that the relative factor becomes zero. So if we had wanted to use the opposite order, it would be related to~(\ref{Map13ef}) by a divergent factor:
\be
e^{(1)}_{II}(z)\overset{\text{opp.}}{=}\eone(z)\Ecal^\half_{h_2}(z)=\lim_{z'\ra z} [\varphi_2(z-z'+h_2)\varphi_2(z-z'-h_2)]^{-\half}~ \Ecal^\half_{h_2}(z') \eone(z) \;.
\ee
The overall $\varphi$ factors for the phase II OPEs do not depend on the ordering. However, as we will see, it is the orderings in~(\ref{Map13ef}) which most simply lead to vanishing of the $\psi$-dependent terms that arise due to the $e-f$ relations.

\subsection{Checking the map} \label{Checks}

We already checked that the map~(\ref{Map13ef}) gives the correct $e^{(3)}-e^{(1)}$ and $f^{(3)}-f^{(1)}$ relations for phase II, given the OPEs of phase I. In this section we will check that the remaining relations between the fermionic fields are also correctly reproduced. We will not show all the possible relations, but some indicative examples which illustrate the general pattern. 

First we need to check that fields belonging to the same node do not acquire any nontrivial bond factors through the map.\footnote{This is the opposite situation to the generalised conifold case, where the map  needs to introduce/remove such bond factors since fermionic adjacent operators are mapped to bosonic ones and vice versa, i.e., the map must create/remove self-lines accordingly.} As an example, consider the $e^{(3)}-e^{(3)}$ relation:
\be
\begin{split}
e_{II}^{(3)}(z) &e_{II}^{(3)}(w)=\ethree(z) \left(\Ecal^\half_{h_1}(z)\ethree(w)\right) \Ecal^\half_{h_1}(w) \\ &\;\;\simeq
-(\varphi_1(u-h_1)\varphi_1(u+h_1))^{-\half} \ethree(w)\left(\ethree(z)\Ecal^\half_{h_1}(w)\right)\Ecal^\half_{h_1}(z)\\
&\;\;\simeq
-(\varphi_1(u\!-\!h_1)\varphi_1(u\!+\!h_1))^{-\half} (\varphi_1(u\!-\!h_1)\varphi_1(u\!+\!h_1))^{\half}\ethree(w)\Ecal^\half_{h_1}(w)\ethree(z)\Ecal^\half_{h_1}(z)\\
&\;\;\simeq -e_{II}^{(3)}(w) e_{II}^{(3)}(z)\;.
\end{split}
\ee
Similarly, we can verify that the $e_{II}^{(1)}-e_{II}^{(1)}$, $f_{II}^{(1)}-f_{II}^{(1)}$ and $f_{II}^{(3)}-f_{II}^{(3)}$ relations do not acquire any nontrivial bond factors, so the respective fields are simply anticommuting.

By the general arguments following~(\ref{Map13initial}), the $e^{(1,3)}-f^{(4)}$ and $e^{(1,3)}-\psi^{(4)}$ OPEs will straightforwardly work out. 
One has to be more careful when considering OPEs of $e_{II}^{(1,3)}(z)$ with $e^{(4)}_{II}(w)$, and similarly $f_{II}^{(1,3)}(z)$ with $f^{(4)}_{II}(w)$. Given~(\ref{Map4}), $e^{(4)}_{II}(w)$ contains an $\ffour(w)$, which does not simply commute with $\Ecal^\half$ due to the $e-f$ relations in~(\ref{QYall}). As an example, let us calculate
\be \label{e1e4}
\begin{split}
  e^{(1)}_{II}(z) e^{(4)}_{II}(w)&=\left(\Ecal^\half_{h_2}(z) \eone(z)\right)\left( \ffour(w)\tpsifour(w)\right)\\
  &\simeq-\Ecal^\half_{h_2}(z)\ffour(w) \left(\varphi_2(z-w) \tpsifour(w) \eone(z)\right)\\
  &\simeq-\varphi_2(z-w)\left(\ffour(w)\tpsifour(w)\right) \left(\Ecal^\half_{h_2}(z)\eone(z)\right)  \\
  &\qquad +\varphi_2(z-w)[\ffour(w),\Ecal^\half_{h_2}(z)]\tpsifour(w)\eone(z)\;.
\end{split}
\ee
The first term gives the required OPE, while to evaluate the second one we need the relation (see Appendix~\ref{AppendixCommutators})
\be
[\ffour(w),\Ecal^\half_{h_2}(z)]=\half \Ecal_{h_2}^{-\half}(z)[\ffour(w),\Ecal_{h_2}(z)]\;. 
\ee
We compute
\be
\begin{split}
  [\ffour(w), &\efour(z+h_2)\efour(z-h_2)]\\
  &=\pb{\ffour(w),\efour(z+h_2)}\efour(z-h_2)-\efour(z+h_2)\pb{\ffour(w),\efour(z-h_2)}\\
  &=\frac{\psifour(w)-\psifour(z+h_2)}{z-w+h_2}\efour(z-h_2)-\efour(z+h_2)\frac{\psifour(w)-\psifour(z-h_2)}{z-w-h_2}\;.
\end{split}
   \ee
However, in~(\ref{e1e4}) this term is multiplied by a $\varphi_2(z-w)=(z-w-h_2)(z-w+h_2)$ factor, which vanishes, up to regular terms, due to the delta functions.\footnote{Alternatively, we could remove this factor by passing the $\tpsi^{(4)}(w)$ back through the $\eone(z)$ field, however the $\psifour-\eone$ relations would bring it back.} We conclude that the delta-function terms vanish and we correctly obtain   
   \be
  e^{(1)}_{II}(z) e^{(4)}_{II}(w)\simeq-\varphi_2(u) e^{(4)}_{II}(w) e^{(1)}_{II}(z)\;.
   \ee
   Similar arguments can be shown to apply for the other OPEs that involve the $e-f$ relations. Let us, for instance, consider the $e^{(3)}-f^{(1)}$ relations, which give
   \be
   e^{(3)}_{II}(z)f^{(1)}_{II}(w)=-f^{(1)}_{II}(w) e^{(3)}_{II}(z)+\ethree(z)\fone(w)[\Ecal_{h_1}^\half(z),\Fcal_{h_2}^\half(w)]\;.
   \ee
Using~(\ref{EhFhstandard}), to leading order we can express the commutator in terms of an $\Ecal(z),\Fcal(w)$ commutator, which will lead to terms containing delta functions of the type
   \be\begin{split}
   &\frac{\psifour(z\!+\!h_1)\!-\!\psifour(w\!+\!h_2)}{z\!-\!w\!+\!h_1\!-\!h_2}\;,\;\frac{\psifour(z\!+\!h_1)\!-\!\psifour(w\!-\!h_2)}{z\!-\!w\!+\!h_1\!+\!h_2}\;,\\
  &\qquad \frac{\psifour(z\!-\!h_1)\!-\!\psifour(w\!+\!h_2)}{z\!-\!w\!-\!h_1\!-\!h_2}\;\;\text{and}\;\; \frac{\psifour(z\!-\!h_1)\!-\!\psifour(w\!-\!h_2)}{z\!-\!w\!-\!h_1\!+\!h_2}\;.\end{split}
   \ee
   We cannot commute the $\ethree(z)$ or $\fone(w)$ on the left through the $\psi^{(4)}$'s depending on the same variable ($z$ or $w$) as that would be equivalent to changing the choice of ordering in~(\ref{Map13ef}), which we wish to keep (as discussed at the end of Section~\ref{MapAdjacent}). However we can use the OPEs with the $\psi$'s depending on the other variable, to write, e.g.,
   \be\begin{split}
   &\ethree(z)\psifour(w\!+\!h_2)\simeq\varphi_1(u\!-\!h_2)\psifour(w\!+\!h_2)\ethree(z)\sim(z\!-\!w\!+\!h_1\!-\!h_2)(z\!-\!w\!-\!h_1\!-\!h_2),\\
   &\fone(w)\psifour(z\!+\!h_1)\simeq\varphi_2(-u\!-\!h_1) \psifour(z\!+\!h_1) \fone(w)\sim(z\!-\!w\!+\!h_1\!+\!h_2)(z\!-\!w\!+\!h_1\!-\!h_2),
   \end{split}\ee
which are both set to zero, up to regular terms, by the first delta function above. Similarly one can see that the remaining three terms vanish, and therefore confirm the $e^{(3)}-f^{(1)}$ relation. The higher terms in~(\ref{EhFhstandard}) contain the same delta functions and can be treated similarly. Checking the $e^{(1)}-f^{(3)}$ relation proceeds in an analogous way, with the main difference being that the $\eone(z),\fthree(w)$ terms will appear to the \emph{right} of the delta functions, as needed in that case to produce vanishing bond factors in the numerator.

\subsection{The $\psi$ fields} \label{SectionPsi}

Turning now to the bosonic $\psi_{II}$ fields for the adjacent nodes, we define
\be  \label{Map13psi}
\boxed{\begin{split}
\psi_{II}^{(1)}(z)&=\Ecal^\half_{h_2}(z)\psione(z)\Fcal^\half_{h_2}(z)\;  ,\\
\psi_{II}^{(3)}(z)&= \Fcal^\half_{h_1}(z)\psithree(z) \Ecal^\half_{h_1}(z)\;,
\end{split}}
\ee
where the $\psi_{II}$ are of course bosonic due to the even numbers of fields appearing in the map. 

It is straightforward to check that all the $\psi-e$, $\psi-f$ and $\psi-\psi$ relations work out correctly. For many of these checks, one needs to consider $\Ecal^\half-\Fcal^\half$ commutators (computed in Appendix~\ref{AppendixCommutators}), which will lead to unwanted delta-function terms. However, these terms always appear with suitable bond factors which are set to zero by the delta functions. Similar arguments are needed in checking that the $e-f$ relations are consistent with the above definition. Let us look at this case in detail. For concreteness, we focus on node 1:
\be \label{eonefone}
\begin{split}
  \{e^{(1)}_{II}(z),f^{(1)}_{II}(w)\}&=\{\Ecal^\half_{h_2}(z)\eone(z),\fone(w)\Fcal^\half_{h_2}(w)\}\\
  &=\Ecal^\half_{h_2}(z)\{\eone(z),\fone(w)\}\Fcal^\half_{h_2}(w)-\fone(w) [\Ecal^\half_{h_2}(z),\Fcal^\half_{h_2}(w)] \eone(z)\;.
\end{split}
\ee
The first term gives (up to regular terms)
\be
\Ecal^\half_{h_2}(z)\left(\frac{\psione(z)-\psione(w)}{z-w}\right)\Fcal^\half_{h_2}(w)=\frac{\psi_{II}^{(1)}(z)-\psi_{II}^{(1)}(w)}{z-w}\;,
\ee
where of course we used that the denominator behaves as a delta function in the large $z,w$ limit. So we have shown consistency with~(\ref{Map13psi}), as  long as the second term vanishes. As mentioned, the commutator appearing there is computed in Appendix~\ref{AppendixCommutators}. To leading order, we have (from~(\ref{EhFhstandard}))
\be\label{EhFhone}
[\Ecal^\half_{h_2}(z),\Fcal^\half_{h_2}(w)]=\frac{1}{4}\Ecal^{-\half}_{h_2}(z)[\Ecal_{h_2}(z),\Fcal_{h_2}(w)]\Fcal^{-\half}_{h_2}(w)+\cdots
\ee
The commutator evaluates to 
\be\label{ExpandedEF}
\begin{split}
  [\Ecal_{h_2}(z),\Fcal_{h_2}(w)]&=[\efour(z+h_2)\efour(z\!-\!h_2),\ffour(w+h_2)\ffour(w\!-\!h_2)]\\
  &=-\efour(z\!+\!h_2)\frac{\psifour(z\!-\!h_2)\!-\!\psifour(w\!+\!h_2)}{z\!-\!w\!-\!2h_2}\ffour(w\!-\!h_2)\!\\
  &\;\;+\!\ffour(w\!+\!h_2)\frac{\psifour(z\!+\!h_2)\!-\!\psifour(w\!-\!h_2)}{z\!-\!w\!+\!2h_2}\efour(z\!-\!h_2)\\
  &\;\;\!-\!\frac{\psifour(z\!+\!h_2)\!-\!\psifour(w\!+\!h_2)}{z\!-\!w}\ffour(w\!-\!h_2)\efour(z\!-\!h_2)\!\\
  &\;\;+\!\efour(z\!+\!h_2)\ffour(w\!+\!h_2)\frac{\psifour(z\!-\!h_2)\!-\!\psifour(w\!-\!h_2)}{z\!-\!w}\;.
\end{split}
\ee
As above, we cannot simply commute the $\eone(z)$ or $\fone(w)$ in~(\ref{eonefone}) through $\psifour$'s in~(\ref{ExpandedEF}) depending on the same $z$ or $w$ variable, as those were originally part of the same $e_{II}$ or $f_{II}$ operator where we made a choice of ordering. However we can pass them through $\psifour$'s depending on the other variable:
\be\begin{split}
\fone(w)\psifour(z\!\pm\! h_2)\simeq \varphi_2(u\!\pm\! h_2)\psifour(z\!\pm\! h_2)\fone(w) \sim (z\!-\!w)(z\!-\!w\mp 2h_2)\;,\\
\psifour(w\!\pm\! h_2)\eone(z)\simeq\varphi_2(u\!\pm\! h_2)\eone(z)\psifour(w\!\pm\! h_2)\sim(z\!-\!w)(z\!-\!w\mp 2h_2)\;,
\end{split}
\ee
and the resulting factors are set to zero by the delta functions. Since the same delta functions appear in the higher terms in~(\ref{EhFhone}), it follows that the second term in~(\ref{eonefone}) vanishes, as required.\footnote{Of course, such statements are always up to regular terms, not affecting the mode expansion for large $z,w$, which we do not keep track of.} The computation for node 3 is essentially identical.

We conclude that~(\ref{Map4}),~(\ref{Map13ef}) and~(\ref{Map13psi}) express the fields of phase II in terms of those of phase I in a way that reproduces all the required OPEs.

\subsection{The inverse map} \label{InverseMap}

Let us now consider the inverse map, which expresses the phase I fields in terms of those of phase II. For node 4 it can be simply read off from~(\ref{Map4}), while the map for the adjacent nodes follows the same pattern as~(\ref{Map13ef}) and~(\ref{Map13psi}), that is: 
\be\begin{split}\label{Mapadjacentinverse}
\eone(z)&\!=\!e^{(1)}_{II}(z) \Ecal^\half_{II,h_2}(z),\;\fone(z)\!=\! \Fcal^\half_{II,h_2}(z)f^{(1)}_{II}(z),\;\psione(z)\!=\!\Fcal^\half_{II,h_2}(z)\psi_{II}^{(1)}(z)\Ecal^\half_{II,h_2}(z)\;,\\
\ethree(z)&\!=\! \Ecal^{\half}_{II,h_1}(z)e^{(3)}_{II}(z),\;\fthree(z)\!=\!f_{II}^{(3)}(z)  \Fcal^{\half}_{II,h_1}(z),\;\psithree(z)\!=\!\Ecal^{\half}_{II,h_1}(z)\psi^{(3)}_{II}(z)\Fcal^{\half}_{II,h_1}(z)\;,
\end{split}
\ee
where now we define $\Ecal_{II,h}(z)=e^{(4)}_{II}(z+h)e^{(4)}_{II}(z-h)$ and $\Fcal_{II,h}(z)=f^{(4)}_{II}(z+h)f^{(4)}_{II}(z-h)$.
Note the reverse order compared to~(\ref{Map13ef}) and~(\ref{Map13psi}), which is related to the inversion of the bond factors compared to phase I. Using the OPEs of phase II,  is straightforward to confirm that
\be
\ethree(z) \eone(w) \sim -\eone(w) \ethree(z)\;,
\ee
as required. All the other OPEs can be seen to work out correctly as well. However, for consistency, it is important to show the above map by explicitly inverting the maps~(\ref{Map13ef}) and~(\ref{Map13psi}). To do this, we need the result~(\ref{EhFhlimit}) stating that $[\Ecal^\half_{h_2}(z),\Fcal^\half_{h_2}(w)]$ diverges as $w\ra z$ as

\be \label{EFcom}
\lim_{w\ra z}[\Ecal^\half_{h_2}(z),\Fcal^\half_{h_2}(w)] =\lim_{u\ra0} \frac{h_2\sqrt{\psifour(z+h_2)\psifour(z-h_2)}}{\sqrt{\varphi_2(u+h_2)\varphi_2(u-h_2)}}\;.
\ee
and similarly for the case with $h_1$. As an example, let us consider the case of $e^{(1)}$. Starting from the direct map, multiplying by $\Fcal^\half_{h_2}$ from the right, and using the node-4 map~(\ref{Map4}) on the left-hand side, we have
\be
\begin{split}
  e_{II}^{(1)}(z)&= \Ecal^\half_{h_2}(z) \eone(z) \;\;\Rightarrow \;\;
  e_{II}^{(1)}(z) \Fcal^\half_{h_2}(w)=\Ecal^\half_{h_2}(z)\Fcal^\half_{h_2}(w) \eone(z) \Rightarrow\\
  e_{II}^{(1)}(z)& \Ecal^\half_{II,h_2}(w)\sqrt{\psit^{(4)}_{II}(w\!+\!h_2)\psit^{(4)}_{II}(w\!-\!h_2)}=\!\left(\Fcal^\half_{h_2}(w)\Ecal^\half_{h_2}(z)+[\Ecal^\half_{h_2}(z),\Fcal^\half_{h_2}(w)]\right) \eone(z)\;.
\end{split}
\ee
Taking the $w\ra z$ limit, the commutator part on the right-hand-side diverges as in~(\ref{EFcom}) and dominates over the first term on the right-hand-side, which can be dropped. Matching this divergence on the left-hand-side requires passing the root-$\psit^{(4)}_{II}\psit^{(4)}_{II}$ term (again defined using the Schwinger parametrisation) to the left of $e^{(1)}_{II}(z)$, which gives the same divergent factor by applying the $e^{(1)}_{II}-\tpsi^{(4)}_{II}$ relations. By the node 4 map (\ref{Map4}), $\psit^{(4)}_{II}=\psifour$, so we can now cancel the root-$\psi$ expressions on both sides to obtain the first relation in~(\ref{Mapadjacentinverse}).\footnote{Up to a factor of $h_2$, which is relevant in ensuring that the $h_2\ra 0$ limit of the map is regular. However, it can be seen that the same factor will arise for the $f$ fields, and its square for the $\psi$ fields, so it can be consistently rescaled away.} Similar arguments can be used to show the other inverse maps.

\subsection{A remark on simpler maps}

One could wonder why one had to define the above maps instead of a much simpler option involving a single field from node 4, which would be more in line with the generalised conifold case~\cite{Bao:2022jhy,Bao:2023kkh}. Since for $\Fzero$ the fields do not change grading, the only possibility would be expressions such as 
\be
e_{II}^{(1)}(z)=\eone(z) \psifour(z_1)\;\;,\;\; e_{II}^{(3)}(z)=\ethree(z) \psifour(z_2)\;,
\ee
where $z_1,z_2$ would be appropriate shifts. However, since the $e^{(1)}\!-\!\psi^{(4)}$ OPEs involve $\varphi_2(u)$ while the $e^{(3)}\!-\!\psi^{(4)}$ OPEs involve $\varphi_1(u)$, it is not possible to construct $\varphi_3(u)$ by using~(\ref{phi3separate}). One could bypass this problem by involving $\psi^{(2)}$, which in the case of $\Fzero$ does not cause any issues, since node 2 is not connected to any further nodes beyond 1 and 3. For instance, one could write
\be
e_{II}^{(1)}(z)=\eone(z) \psifour(z+h_2)\;\;,\;\; e_{II}^{(3)}(z)=\ethree(z) \tpsitwo(z+h_2)\;,
\ee
which indeed gives
\be
e_{II}^{(3)}(z)e_{II}^{(1)}(w)=-\varphi_1(u+h_2)\varphi_1(u-h_2)~e_{II}^{(1)}(w)e_{II}^{(3)}(z)=-\varphi_3(u)~e_{II}^{(1)}(w)e_{II}^{(3)}(z)\;,
\ee
using the $\psi^{(2)}\!-\!e^{(1)}$ relations. However, one eventually runs into the problem that the $\psi$ fields treat $e$ and $f$ oppositely in terms of bond factors. So if one were to define
\be \label{falt}
f_{II}^{(1)}(z)=\fone(z) \psi^{(4)}(z-h_2)\;\;,\;\; f_{II}^{(3)}(z)=\fthree(z) \tpsi^{(2)}(z-h_2)\;,
\ee
which give the correct  $\varphi_3^{-1}$ in the $f^{(3)}\!-\!f^{(1)}$ relations, then one also satisfies the $e^{(3)}\!-\!f^{(1)}$ and $e^{(1)}\!-\!f^{(3)}$ relations but obtains unwanted bond factors in the $e^{(3)}\!-\!f^{(3)}$ and $e^{(1)}\!-\!f^{(1)}$ relations. Swapping $\{\psi^{(4)},\tpsi^{(2)}\}\ra \{\tpsi^{(2)},\psi^{(4)}\}$ in~(\ref{falt}) would still give the correct $f^{(3)}\!-\!f^{(1)}$ relation but spoil the remaining ones. Also, of course, maps involving only $\psi$'s for the fermionic fields would imply that $\psi^{(1,3)}_{II}$ would purely depend on $\psi_I$ fields, but that cannot reproduce the $e_{II}-\psi_{II}$ and $f_{II}-\psi_{II}$ relations as the $\psi$'s commute among themselves. Therefore, it appears necessary to introduce the square root maps involving the fermion modes, as in~(\ref{Map13ef}) and~(\ref{Map13psi}).

\section{Discussion}

In this work we provided an explicit isomorphism between the quiver Yangians corresponding to the two Seiberg dual phases of $\Fzero$. The map is slightly exotic in that it involves a square root of a bilinear operator in the fermionic fields of the node being dualised. Given some care in working with such operators, we were able to show that all the OPE relations of phase II can be obtained in terms of those of phase I, and vice versa. A more rigorous approach, perhaps along the lines of expressing the fermion fields as boson-fermion composites (as in the free-field descriptions~\cite{Gaberdiel:2017hcn}) would certainly be desirable and is left for future work.    

The square root nature of the map clearly complicates the mode expansion, and mapping the modes of phase II to those of phase I is not straightforward within our approach. Therefore, it would be relevant to attempt to construct such a map directly at the level of the modes (of course assuming that such a local map exists, which might not be the case), and, if successful, eventually build up the map for the fields and compare with our map. Of course, we do not claim uniqueness and it might be possible to find alternative maps which are more amenable to be expanded in modes. On the other hand, it is possible that the non-analyticity we find is a feature, rather than a bug. As mentioned in the introduction, the crystal representations differ significantly across chambers related by Seiberg duality, even in cases where the quivers themselves are invariant, see \cite{Galakhov:2021xum, Galakhov:2024foa} in the quiver Yangian context. This implies that the modes (which build up the crystals in each chamber), as well as other algebraic structures such as the coproduct, can be expected to map in a non-trivial way. It will be important to establish whether the square-root map we find for the currents is consistent with this intricate behaviour. 

As our focus was purely on the quadratic relations, we did not consider how the higher order/Serre relations are mapped between the two phases. In general the Serre relations for quivers with compact four-cycles are poorly understood. However, a proposal for deriving the Serre relations for such chiral quivers is given in~\cite{Galakhov:2021vbo}, and the resulting relations for phase I of $\Fzero$ are explicitly worked out there. Applying our map to correctly obtain the Serre relations for phase II could provide an important check of this proposal. 

It would be interesting to extend our results to other theories with compact four-cycles, such as the case of the degree-3 del Pezzo quiver $\mathrm{dP}_3$, where there are four distinct phases (see, e.g.,~\cite{Feng:2002zw,Franco:2005rj}) and therefore one would expect to find four isomorphic quiver Yangians. This case is substantially different to the $\Fzero$ case we considered in this work, as most arrows are single (leading to linear bond factors) and also Seiberg dualising does not just invert all arrows involving the dualised node, but also modifies the charge assignments in a non-trivial way. The eventual goal would be to come up with a simple graphical procedure for dualising the algebras, perhaps at the level of the bipartite graph (similarly to urban renewal), which would cover all quivers.\footnote{To this end, it would also be relevant to revisit the isomorphism for the generalised conifold cases~\cite{BezerraMukhin19b,Bao:2022jhy,Bao:2023kkh}, and express it fully in terms of the fields.}

The quiver Yangians have been generalised to toroidal and elliptic algebras~\cite{Noshita:2021ldl,Galakhov:2021vbo} and it would be natural to try to extend our results to those cases as well. Since the product nature of the map lends itself to a direct generalisation of relations like~(\ref{phi3map}) to products of trigonometric and elliptic functions, we expect this extension to be relatively straightforward.\footnote{In fact, it might be expected to work better, since in those cases the $e-f$ relations contain true delta functions so one doesn't have to work up to regular terms.} 

 Finally, quiver Yangians are closely related to other algebraic structures (such as Yang-Baxter algebras and $\Wcal$-algebras~\cite{Prochazka:2015deb, Gaberdiel:2017dbk,Gaberdiel:2017hcn, Bao:2022fpk}), which are still being explored, and we expect that a better understanding of how quiver Yangians transform under Seiberg duality will have implications for those relationships as well.

\section*{Acknowledgements}

The authors are grateful to Jiakang Bao, Dmitry Galakhov and Wei Li for useful correspondence. VJ and DN are supported by the South African Research Chairs Initiative of the Department of Science, Technology, and Innovation (DSTI) and the National Research Foundation (NRF), grant 78554. KZ wishes to thank the theory group at DESY, Hamburg and the Cluster of Excellence: Quantum Universe at the University of Hamburg, as well as the Niels Bohr Institute, Copenhagen, for their hospitality during the later stages of this work.

\appendix 
 
\section{Quadratic relations for the $\Fzero$ quiver Yangian} \label{AppendixQuadratic}
  
For easy reference, in this appendix we reproduce the quadratic relations of the two phases of $\Fzero$. In the following $u=z-w$, and since the bond factors in this case are even, we do not distinguish between $\varphi_i(u)$ and $\varphi_i(-u)$. For both phases we have:
\be
\begin{split}
  & \psi^{(a)}(z)\psi^{(b)}(w)=\psi^{(b)}(w)\psi^{(a)}(z)\;,\\
  & \psi^{(a)}(z) e^{(a)}(w)\simeq e^{(a)}(w)\psi^{(a)}(z)\;,\\
 & \psi^{(a)}(z) f^{(a)}(w)\simeq f^{(a)}(w)\psi^{(a)}(z)\;,\\
  &e^{(a)}(z) e^{(a)}(w)\sim-e^{(a)}(w)e^{(a)}(z)\;,\\
  &f^{(a)}(z) f^{(a)}(w)\sim-f^{(a)}(w)f^{(a)}(z)\;,\\
  \end{split}
    \ee
and 
        \be
  \{e^{(a)}(z),f^{(b)}(w)\}=-\delta^{a,b}\frac{\psi^{(a)}(z)-\psi^{(b)}(w)}{z-w}\;.
\ee
As we are dualising node 4, node 2 remains unaffected, as well as the arrows linking it to nodes 1 and 3. So the following relations are common for the two phases:
\be
    \begin{split}
  &\psi^{(1)}(z)e^{(2)}(w)\simeq \varphi_1(u)~ e^{(2)}(w)\psi^{(1)}(z)\;,\\
  & \psi^{(2)}(z)e^{(1)}(w)\simeq \varphi_1(u)^{-1}~ e^{(1)}(w)\psi^{(2)}(z)\;,\\
&\psi^{(1)}(z)f^{(2)}(w)\simeq \varphi_1(u)^{-1} ~f^{(2)}(w)\psi^{(1)}(z)\;,\\
  & \psi^{(2)}(z)f^{(1)}(w)\simeq \varphi_1(u)~ f^{(1)}(w)\psi^{(2)}(z)\;,\\
  &e^{(1)}(z)e^{(2)}(w)\sim-\varphi_1(u) ~e^{(2)}(w) e^{(1)}(z)\;,\\
 &f^{(1)}(z)f^{(2)}(w)\sim-\varphi_1(u)^{-1}~ f^{(2)}(w) f^{(1)}(z)\;,\\
 \end{split} 
  \ee
and
  \be
  \begin{split}
  &\psi^{(2)}(z)e^{(3)}(w)\simeq \varphi_2(u) ~e^{(3)}(w)\psi^{(2)}(z)\;,\\
  & \psi^{(3)}(z)e^{(2)}(w)\simeq \varphi_2(u)^{-1}~ e^{(2)}(w)\psi^{(3)}(z)\;,\\
&\psi^{(2)}(z)f^{(3)}(w)\simeq \varphi_2(u)^{-1} ~f^{(3)}(w)\psi^{(2)}(z)\;,\\
  & \psi^{(3)}(z)f^{(2)}(w)\simeq \varphi_2(u)~ f^{(2)}(w)\psi^{(3)}(z)\;,\\
    &e^{(2)}(z)e^{(3)}(w)\sim-\varphi_2(u)~e^{(3)}(w) e^{(2)}(z)\;,\\
 &f^{(2)}(z)f^{(3)}(w)\sim-\varphi_2(u)^{-1} ~f^{(3)}(w) f^{(2)}(z)\;,\\
 \end{split} 
  \ee
as well as
  \be
  \begin{split}
  &\psi^{(4)}(z)e^{(2)}(w)\simeq e^{(2)}(w)\psi^{(4)}(z)\;,\\
  & \psi^{(2)}(z)e^{(4)}(w)\simeq  e^{(4)}(w)\psi^{(2)}(z)\;,\\
&\psi^{(4)}(z)f^{(2)}(w)\simeq  f^{(2)}(w)\psi^{(4)}(z)\;,\\
  & \psi^{(2)}(z)f^{(4)}(w)\simeq f^{(4)}(w)\psi^{(2)}(z)\;,\\
  &e^{(4)}(z)e^{(2)}(w)\sim- e^{(2)}(w) e^{(4)}(z)\;,\\
 &f^{(4)}(z)f^{(2)}(w)\sim- f^{(2)}(w) f^{(4)}(z)\;.\\
 \end{split} 
  \ee
 The following relations differ between the two phases:
  
\subsection{Phase I} \label{AppPhaseI}

  \be \label{PhaseI34}
  \begin{split}
  &\psi^{(3)}_I(z)e_I^{(4)}(w)\simeq \varphi_1(u) ~e_I^{(4)}(w)\psi^{(3)}_I(z)\;,\\
  & \psi^{(4)}_I(z)e^{(3)}_I(w)\simeq \varphi_1(u)^{-1}~ e^{(3)}_I(w)\psi^{(4)}_I(z)\;,\\
    &\psi^{(3)}_I(z)f^{(4)}_I(w)\simeq \varphi_1(u)^{-1}~f^{(4)}_I(w)\psi^{(3)}_I(z)\;,\\
  & \psi^{(4)}_I(z)f^{(3)}_I(w)\simeq \varphi_1(u) ~f^{(3)}_I(w)\psi^{(4)}_I(z)\;,\\
  &e^{(3)}_I(z)e^{(4)}_I(w)\sim-\varphi_1(u)~ e^{(4)}_I(w) e^{(3)}_I(z)\;,\\
 &f^{(3)}_I(z)f^{(4)}_I(w)\sim-\varphi_1(u)^{-1} ~f^{(4)}_I(w) f^{(3)}_I(z)\;,\\
 \end{split} 
  \ee

    \be \label{PhaseI41}
  \begin{split}
  &\psi^{(4)}_I(z)e^{(1)}_I(w)\simeq \varphi_2(u)~ e^{(1)}_I(w)\psi^{(4)}_I(z)\;,\\
  & \psi^{(1)}_I(z)e^{(4)}_I(w)\simeq \varphi_2(u)^{-1}~ e^{(4)}_I(w)\psi^{(1)}_I(z)\;,\\
&\psi^{(4)}_I(z)f^{(1)}_I(w)\simeq \varphi_2(u)^{-1}~ f^{(1)}_I(w)\psi^{(4)}_I(z)\;,\\
  & \psi^{(1)}_I(z)f^{(4)}_I(w)\simeq \varphi_2(u)~ f^{(4)}_I(w)\psi^{(1)}_I(z)\;,\\
  &e^{(4)}_I(z)e^{(1)}_I(w)\sim-\varphi_2(u)~ e^{(1)}_I(w) e^{(4)}_I(z)\;,\\
 &f^{(4)}_I(z)f^{(1)}_I(w)\sim-\varphi_2(u)^{-1} ~f^{(1)}_I(w) f^{(4)}_I(z)\;,\\
 \end{split} 
  \ee

      \be \label{PhaseI31}
  \begin{split}
  &\psi^{(3)}_I(z)e^{(1)}_I(w)\simeq e^{(1)}_I(w)\psi^{(3)}_I(z)\;,\\
  & \psi^{(1)}_I(z)e^{(3)}_I(w)\simeq  e^{(3)}_I(w)\psi^{(1)}_I(z)\;,\\
&\psi^{(3)}_I(z)f^{(1)}_I(w)\simeq  f^{(1)}_I(w)\psi^{(3)}_I(z)\;,\\
  & \psi^{(1)}_I(z)f^{(3)}_I(w)\simeq f^{(3)}_I(w)\psi^{(1)}_I(z)\;,\\
  &e^{(3)}_I(z)e^{(1)}_I(w)\sim- e^{(1)}_I(w) e^{(3)}_I(z)\;,\\
 &f^{(3)}_I(z)f^{(1)}_I(w)\sim- f^{(1)}_I(w) f^{(3)}_I(z)\;.\\
 \end{split} 
  \ee

\subsection{Phase II} \label{AppPhaseII}

  \be \label{PhaseII34}
  \begin{split}
  &\psi^{(3)}_{II}(z)e^{(4)}_{II}(w)\simeq \varphi_1(u)^{-1}~ e^{(4)}_{II}(w)\psi^{(3)}_{II}(z)\;,\\
  & \psi^{(4)}_{II}(z)e^{(3)}_{II}(w)\simeq \varphi_1(u) ~e^{(3)}_{II}(w)\psi^{(4)}_{II}(z)\;,\\
&\psi^{(3)}_{II}(z)f^{(4)}_{II}(w)\simeq \varphi_1(u) ~f^{(4)}_{II}(w)\psi^{(3)}_{II}(z)\;,\\
  & \psi^{(4)}_{II}(z)f^{(3)}_{II}(w)\simeq \varphi_1(u)^{-1}~ f^{(3)}_{II}(w)\psi^{(4)}_{II}(z)\;,\\
  &e^{(3)}_{II}(z)e^{(4)}_{II}(w)\sim-\varphi_1(u)^{-1}~ e^{(4)}_{II}(w) e^{(3)}_{II}(z)\;,\\
 &f^{(3)}_{II}(z)f^{(4)}_{II}(w)\sim-\varphi_1(u) ~f^{(4)}_{II}(w) f^{(3)}_{II}(z)\;,\\
 \end{split} 
  \ee

    \be \label{PhaseII41}
  \begin{split}
  &\psi^{(4)}_{II}(z)e^{(1)}_{II}(w)\simeq \varphi_2(u)^{-1}~ e^{(1)}_{II}(w)\psi^{(4)}_{II}(z)\;,\\
  & \psi^{(1)}_{II}(z)e^{(4)}_{II}(w)\simeq \varphi_2(u) ~e^{(4)}_{II}(w)\psi^{(1)}_{II}(z)\;,\\
&\psi^{(4)}_{II}(z)f^{(1)}_{II}(w)\simeq \varphi_2(u) ~f^{(1)}_{II}(w)\psi^{(4)}_{II}(z)\;,\\
  & \psi^{(1)}_{II}(z)f^{(4)}_{II}(w)\simeq \varphi_2(u)^{-1}~ f^{(4)}_{II}(w)\psi^{(1)}_{II}(z)\;,\\
  &e^{(4)}_{II}(z)e^{(1)}_{II}(w)\sim-\varphi_2(u)^{-1} ~e^{(1)}_{II}(w) e^{(4)}_{II}(z)\;,\\
 &f^{(4)}_{II}(z)f^{(1)}_{II}(w)\sim-\varphi_2(u)~ f^{(1)}_{II}(w) f^{(4)}_{II}(z)\;,\\
 \end{split} 
  \ee

      \be \label{PhaseII31}
  \begin{split}
  &\psi^{(3)}_{II}(z)e^{(1)}_{II}(w)\simeq \varphi_3(u)~ e^{(1)}_{II}(w)\psi^{(3)}_{II}(z)\;,\\
  & \psi^{(1)}_{II}(z)e^{(3)}_{II}(w)\simeq \varphi_3(u)^{-1}~ e^{(3)}_{II}(w)\psi^{(1)}_{II}(z)\;,\\
&\psi^{(3)}_{II}(z)f^{(1)}_{II}(w)\simeq \varphi_3(u)^{-1} ~f^{(1)}_{II}(w)\psi^{(3)}_{II}(z)\;,\\
  & \psi^{(1)}_{II}(z)f^{(3)}_{II}(w)\simeq \varphi_3(u) ~f^{(3)}_{II}(w)\psi^{(1)}_{II}(z)\;,\\
  &e^{(3)}_{II}(z)e^{(1)}_{II}(w)\sim- \varphi_3(u) ~e^{(1)}_{II}(w) e^{(3)}_{II}(z)\;,\\
 &f^{(3)}_{II}(z)f^{(1)}_{II}(w)\sim-\varphi_3(u)^{-1} ~f^{(1)}_{II}(w) f^{(3)}_{II}(z)\;.\\
 \end{split} 
  \ee

  \section{Working with the $\Ecal^\half$ and $\Fcal^\half$ operators} \label{AppendixCommutators}

  In this appendix we derive some useful formulas for computations with the $\Ecal^\half$ and $\Fcal^\half$ operators. To lighten the notation, we will not indicate the node (which is always 4 for all fields in this appendix), nor the $h_1$ or $h_2$ subscripts, which can be easily reinstated. Recall that half-integer powers of operators can be defined through the Schwinger parametrisation, 
  \be \label{Schwinger}
  \mathcal{O}^{\frac{n}{2}}=-\frac{1}{2\Gamma(-\frac{n}{2})}\oint_C \diff s~ s^{-\frac{n}{2}-1} e^{-s\mathcal{O}}\;,
  \ee
with $C$ being the Hankel contour going around $\mathbb{R}_+$ in the positive direction. We will not need to evaluate any contour integrals, we will rather be applying this formula to compute commutators of $\mathcal{O}^\frac{n}2$ with other operators via expanding the exponential, commuting with the required operator and re-exponentiating. In doing this we will be ignoring the bifermion nature of $\Ecal$ and $\Fcal$. More precisely, we will implicitly assume a point-splitting prescription allowing us to define the powers $\Ecal^n, \Fcal^n$. This will allow us to obtain general formulas, as if $\Ecal$, $\Fcal$ were true bosons, which we will take as valid for our case as well. In particular, some final results will only be in terms of the bosonic $\psi$ fields, at which point we can safely do away with the point splitting. We take the success of the map as an indication that further work can make our manipulations fully consistent. 

  As discussed, OPEs of $\Ecal^\half$ and $\Fcal^\half$ with any fields from nodes other than 4 simply produce the square roots of the corresponding bond factors, due to the homogeneous nature of these OPEs. The only complications arise when considering OPEs with other fields of node 4, as the $e-f$ relations contain a non-homogeneous term involving $\psi$'s, and we will focus on such terms. Let us start by considering the commutator of $\Ecal^\half$ with $f^{(4)}$. We have
  \be\begin{split}
           [f(w)&,~\Ecal^\half(z)]=\frac{1}{4\sqrt{\pi}}\oint\diff s\; s^{-\frac32}\left([f(w),1]-s[f(w),\Ecal(z)]+\frac{s^2}{2}[f(w),\Ecal^2(z)]+\cdots\right)\\
           &=-\frac{1}{4\sqrt{\pi}}\oint\diff s\; s^{-\frac12}\big([f(w),\Ecal(z)]\!-\!\frac{s}{2}\big(2 \Ecal(z) [f(w),\Ecal(z)]-[\Ecal(z),[f(w),\Ecal(z)]]\big)+\cdots\big)\;,
  \end{split}
  \ee
  where we used that $[A,B^2]=B[A,B]+[A,B]B=2B[A,B]-[B,[A,B]]$. However, the nested commutator term is zero since $[f(w),\Ecal(z)]$ does not contain any $f$'s. Therefore we can simply re-exponentiate the $\Ecal$'s to obtain
  \be
  [f(w),~\Ecal^\half(z)]=-\frac{1}{4\sqrt{\pi}}\left(\oint\diff s\; s^{-\frac{1}{2}} e^{-s\Ecal(z)} \right)[f(w),\Ecal(z)]
=\half\Ecal^{-\half}(z)[f(w),\Ecal(z)]\;,
  \ee
where we note that the relevant power in~(\ref{Schwinger}) is now $\frac{n}{2}=-\half$. This is the formula used in Section~\ref{Checks} as part of checking the $e^{(1)}-f^{(4)}$ relations. 

The above result is standard as it is similar to the case of operators with constant commutator, and is really just expressing the derivation property of the commutator. However, we will also require the commutators of $\Ecal^\half$ with $\Fcal^\half$, where the double-commutator terms do contribute and more care is required. To see this, consider first
\be
\begin{split}
[\Ecal(z),\Fcal^\half(w)]&=\frac{1}{4\sqrt{\pi}}\oint\diff s\; s^{-\frac{3}{2}}\left([\Ecal,1]-s [\Ecal,\Fcal]+\frac{s^2}{2}\left(\Fcal [\Ecal,\Fcal]
+[\Ecal,\Fcal]\Fcal\right)+\cdots \right)\\ 
&=-\frac{1}{4\sqrt{\pi}}\oint\diff s\; s^{-\half} \left([\Ecal,\Fcal]-s[\Ecal,\Fcal]\Fcal-\frac{s}2 [\Fcal,[\Ecal,\Fcal]]+\cdots\right)\;,
\end{split}
\ee
where we have suppressed the $z,w$ dependence on the right-hand side. 
As we will see below, the nested commutator term is not zero, although further commutators with $\Fcal$ do vanish. We can then write in general
\be
   [\Ecal,\Fcal^k]=k[\Ecal,\Fcal]\Fcal^{k-1}+\frac{k(k-1)}{2}[\Fcal,[\Ecal,\Fcal]]\Fcal^{k-2}\;.
\ee
From this we can easily check that the generic $1/k![\Ecal,\Fcal^k]$ terms will, after moving all $\Fcal$'s to the right, produce a $\half[\Fcal,[\Ecal,\Fcal]]$ nested term, times $1/(k-2)! \Fcal^{k-2}$, with the sum starting from $k=2$. So we can re-exponentiate to find
\be \label{EFh}\begin{split}
  [\Ecal(z),\Fcal^\half(w)]&=-\frac{1}{4\sqrt\pi}[\Ecal,\Fcal]\oint\diff s\; s^{-\half} e^{-s\Fcal}+\frac{1}{8\sqrt\pi}[\Fcal,[\Ecal,\Fcal]] \oint \diff s\; s^\half e^{-s\Fcal}\\
&=
  \half [\Ecal(z),\Fcal(w)]\Fcal^{-\half}(w) -\frac{1}{8} [\Fcal(w),[\Ecal(z),\Fcal(w)]]\Fcal^{-\frac32}(w)\;.
\end{split}   \ee
We will also need the analogous formula for general half-integer powers, which works out to
\be\label{generalEF}
   [\Ecal(z),\Fcal^{\frac{n}{2}}(w)]=\frac{n}2[\Ecal(z),\Fcal(w)]\Fcal^{\frac{n}{2}-1}(w)+\frac{n(n-2)}{8}[\Fcal(w),[\Ecal(z),\Fcal(w)]]\Fcal^{\frac{n}{2}-2}(w)\;,
   \ee
where the coefficient of each term comes from $-\Gamma(-n/2+1)/\Gamma(-n/2)$ and $\half\Gamma(-n/2+2)/\Gamma(-n/2)$, respectively. 
Introducing the compact notation $e_\pm\!=\!e^{(4)}_I(z\pm h)$, $f_\pm\!=\!f^{(4)}_I(z\pm h)$ and $\psi_{\pm\pm}\!=\!\frac{\psi^{(4)}(z\pm h)-\psi^{(4)}(w\pm h)}{(z\pm h)-(w \pm h)}$, we compute   
  \be \label{efpsi}
  \begin{split}
    [\Ecal(z),\Fcal(w)]&=[e_+e_-,f_+f_-]\\
    &=e_+\pb{e_-,f_+}f_-\!-\!f_+\pb{e_+,f_-}e_-\!+\!\pb{e_+,f_+}f_-e_-\!-\!e_+f_+\pb{e_-,f_-}\\
    &=-e_+f_-\psi_{-+}+f_+e_-\psi_{+-}-f_-e_-\psi_{++}+e_+f_+\psi_{--}\;,\\
    &=(ef ~\text{terms})+\Delta\psi=(fe ~\text{terms}-\Delta\psi)\;,
\end{split}
  \ee
  where we have defined
  \be
  \begin{split}
  \Delta\psi=\psi_{++}\psi_{--}\!-\!\psi_{+-}\psi_{-+}
  &=\frac{(\psi(z\!+\!h)\!-\!\psi(w\!+\!h))(\psi(z\!-\!h)\!-\!\psi(w\!-\!h))}{(z-w)^2}\!\\
  &\qquad-\frac{(\psi(z\!+\!h)\!-\!\psi(w\!-\!h))(\psi(z-h)\!-\!\psi(w+h))}{(z-w)^2-4h^2}\;.
\end{split}
  \ee
 As mentioned, the nested commutator does not vanish,
  \be\begin{split}
           &[\Fcal(w),~[\Ecal(z),\Fcal(w)]]\\
           &=-f_+\pb{f_-,e_+}f_-\psi_{-+}\!-\!f_+\pb{f_+,e_-}f_-\psi_{+-}\!+\!f_+f_-\pb{f_-,e_-}e_-\psi_{++}\!-\!f_-\pb{f_+,e_+}f_+\psi_{--}\\
           &=+f_+f_-\psi_{+-}\psi_{-+}+f_+f_-\psi_{+-}\psi_{-+}-f_+f_-\psi_{--}\psi_{++}+f_-f_+\psi_{++}\psi_{--}\\
           &=-2\Fcal \Delta\psi\;.
  \end{split}
  \ee
In the above we have neglected  $(f_+)^2,(f_-)^2$ terms as they will not contribute in the coincident limit. One sees that taking additional commutators with $\Fcal$ does give zero. Similarly, one can show that $[\Ecal,[\Ecal,\Fcal]]=2\Ecal\Delta\psi$.  
We are now ready to consider
   \be \label{EhFhfull}\begin{split}
     [\Ecal^\half(z),\Fcal^\half(w)]&=\frac{1}{4\sqrt{\pi}}\oint\diff s\; s^{-\frac32}\left([1,\Fcal^\half]\!-\!s[\Ecal,\Fcal^\half]\!+\!\frac{s^2}{2}[\Ecal^2,\Fcal^\half]\right)\\
&=-\frac{1}{4\sqrt{\pi}}\oint\diff s\; s^{-\half}\left([\Ecal,\Fcal^\half]-\frac{s}{2}\left(2\Ecal [\Ecal,\Fcal^\half]\!-\![\Ecal,[\Ecal,\Fcal^\half]]\right)+\cdots \right)\;.
    \end{split}\ee
    Expanding each term according to~(\ref{EFh}), moving all $\Ecal$s to the left, and re-exponentiating, we find a standard term of the type
    \be \label{EhFhstandard}
        [\Ecal^\half(z),\Fcal^\half(w)]=\frac{1}{4}\Ecal^{-\half}(z)[\Ecal(z),\Fcal(w)]\Fcal^{-\half}(w)+\cdots
       \ee
This was used in Section~\ref{SectionPsi}, in showing consistency of the $e-f$ relations with the definition of the $\psi$ field. For some other checks, it could be more convenient to chose a different ordering of the $\Ecal$ and $\Fcal$s, for instance.
    \be \label{EhFhalt}
        [\Ecal^\half(z),\Fcal^\half(w)]=\frac{1}{4}\Fcal^{-\half}(w)[\Ecal(z),\Fcal(w)]\Ecal^{-\half}(z)+\cdots
       \ee
where the higher-order terms will be different but will still include powers of $\Delta\psi$ and therefore the same delta functions. It is perhaps less obvious that the root-$\Ecal$, root-$\Fcal$ commutator includes terms that contain purely powers of $\Delta\psi$. A heuristic way to extract these terms is to apply the Schwinger parametrisation to both operators at once: 
\be\begin{split}
         [\Ecal^\half(z),\Fcal^\half(w)]&\sim\oint\diff s~s^{-\frac{3}2}\oint\diff t~ t^{-\frac{3}2} [e^{-s\Ecal},e^{-t\Fcal}]\\
           &=\oint\oint\diff s\; \diff t\; (st)^{-\frac{3}2} \left(st[\Ecal,\Fcal]+\frac{s^2t^2}{4}[\Ecal^2,\Fcal^2]+\cdots\right)\\
           &=\oint\oint \diff s\diff t\; (st)^{-\frac{3}{2}} (st\Delta\psi)\left(1-\frac{st}2 \Delta\psi+\frac{s^2t^2}{6}\Delta\psi^2 +\cdots\right)\;,
\end{split}
\ee
where we focused only on terms with equal numbers of $\Ecal$'s and $\Fcal$'s and extracted the double-commutator terms by moving $\Fcal$'s to the left, e.g.,
\be\begin{split}
[\Ecal^2,\Fcal^2]&=\Ecal[\Ecal,\Fcal]\Fcal+\Ecal\Fcal[\Ecal,\Fcal]+\Fcal[\Ecal,\Fcal]\Ecal+[\Ecal,\Fcal]\Fcal\Ecal\\
&= (-2\Delta\psi)\left(\Ecal\Fcal+\Fcal\Ecal\right)+\cdots=-2\Delta\psi([\Ecal,\Fcal]+2\Fcal\Ecal)+\cdots=2(\Delta\psi)^2+\cdots\;,
\end{split}\ee
where the dots indicate terms of lower order in $\Delta\psi$. We use~(\ref{efpsi}) with the minus, since we are moving $f$'s to the left. The higher order terms give factors of $k!$ cancelling one of the $k!$ terms in the denominator. Changing integration variables as, for instance, $s=r^\half u$, $t=r^\half/u$, one obtains, up to constants and the integral in $u$,
\be\begin{split}
         [\Ecal^\half(z),\Fcal^\half(w)]&\sim \oint\diff r\; r^{-\frac32} (r\Delta\psi)\left(1-\frac{r}2\Delta\psi+\frac{r^2}{6}\Delta\psi^2+\cdots\right)\;\\
         &\sim\oint\diff r\; r^{-\frac32} (r\Delta\psi)\frac{e^{-r\Delta\psi}-1}{r\Delta\psi}+\cdots\sim\sqrt{\Delta\psi}-\oint\diff r\; r^{-\frac{3}{2}}+\cdots\;.
         \end{split}
\ee
Assuming an appropriate regularisation of the last (field-independent) integral, we find that $[\Ecal^\half(z),\Fcal^\half(w)]$ includes a term proportional to $\sqrt{\Delta\psi}$.

An alternative way to obtain this result is to consider the $(\Delta\psi)^k$ contributions at order $s^{k-\frac32}$ in~(\ref{EhFhfull}), with $k\geq 1$. We can focus on terms with an $\Fcal^{\half-k}$ to the left of the commutators that will give the $(\Delta\psi)^k$, and ignore terms without this structure. For this we write
\be
[\Ecal^k,\Fcal^\half]=[\Ecal^k,\Fcal^{\half-k}\Fcal^k]=\Fcal^{\half-k}[\Ecal^k,\Fcal^k]+[\Ecal^k,\Fcal^{\half-k}]\Fcal^k\;.
\ee
As above, we then express $[\Ecal^k,\Fcal^k]=k(-\Delta\psi)^k+\cdots$, which in~(\ref{EhFhfull}) exponentiates as 
\be
  -\frac{1}{4\sqrt{\pi}}\Delta\psi\Fcal^{-\half}\oint\diff s\; s^{-\half} e^{-s\Delta\psi \Fcal^{-1}}=\half\Delta\psi\Fcal^{-\half}\left(\Delta\psi\Fcal^{-1}\right)^{-\half}=\half\sqrt{\Delta\psi} \;.  \ee
So we have arrived at the slightly more precise expression
  \be
     [\Ecal^\half(z),\Fcal^\half(w)]= \half\sqrt{\Delta\psi} +\cdots\;.
     \ee
We stress that these manipulations hide (potentially divergent) contributions which we have not considered and would need to be regularised carefully. However, we expect that the outcome, i.e., that the root-$\Ecal$, root-$\Fcal$ commutator includes a root-$\Delta\psi$ term, is robust. The reason for isolating this factor is that it is divergent in the $w\ra z$ limit, while the terms including lower or no powers of $\Delta\psi$ in~(\ref{EhFhfull}) will be less divergent. The scaling of $\Delta\psi$ can be seen, e.g., by working in the mode expansion for $\psi$, where it is straightforward to check that in the large $z$, $w\ra z$ limit, we have\footnote{To see this result, recall that for large $z,w$, the delta function terms expand as $(\psi(z)-\psi(w))/(z-w)=\sum_k(z^{-k-1}\!-\!w^{-k-1})/(z-w)\psi_k=\sum_k(z^{-k}w^{-1}+\cdots+w^{-k}z^{-1})\psi_k=\sum_k\sum_mz^{-m}w^{-n}\psi_{k+m}$. Applying this to the different components of $\Delta\psi$ straightforwardly gives the result.}
     \be
    (z-w)^2((z-w)^2-4h^2)~ \Delta\psi \;\longrightarrow \; 4 h^2 \psi(z+h)\psi(z-h)+O(1/z)\;,
     \ee
     or in other words
     \be \label{EhFhlimit}
    \lim_{w\ra z} [\Ecal^\half(z),\Fcal^\half(w)]=\lim_{u\ra0} \frac{h\sqrt{\psi(z+h)\psi(z-h)}}{\sqrt{\varphi(u+h)\varphi(u-h)}} \;+\;\text{less divergent terms}\;,
\ee
 where $\varphi$ denotes either $\varphi_1$ or $\varphi_2$ depending on whether the charge is $h_1$ or $h_2$.  We see the appearance of a (divergent) inverse bond factor which would normally be associated to commuting node 4 fields with either node 1 or node 3 fields, despite the fact that our computation was purely in node 4. As shown in section~\ref{InverseMap}, the result~(\ref{EhFhlimit}) is crucial for being able to invert the map and express the phase I fields in terms of those of phase II.

\bibliographystyle{utphys}
\bibliography{quiveryangian}

\end{document}